\documentclass[oneside,11pt]{article}

\usepackage[utf8]{inputenc}
\usepackage[T1]{fontenc}

\usepackage{amsthm}
\usepackage{amsmath}
\usepackage{amsfonts}
\usepackage{amssymb}

\usepackage{mathtools}

\usepackage{graphicx} 
\usepackage{float}
\usepackage{booktabs}
\usepackage{multirow}
\usepackage{rotating}
\usepackage{array}

\usepackage{breakcites}

\usepackage{enumitem}

\usepackage[top=1.5cm,bottom=2cm,right=2.5cm,left=2.5cm]{geometry}
\usepackage{titling}

\usepackage{natbib}

\usepackage{ifplatform}

\ifwindows
\fi

\ifwindows
	\usepackage{bbm}
\else
	\iflinux
		\usepackage{bbm}
	\else
		\usepackage{bbm}
	\fi
\fi

\newcommand{\lp}{\left(}
\newcommand{\rp}{\right)}
\newcommand{\lc}{\left[}
\newcommand{\rc}{\right]}
\newcommand{\lb}{\left\{}
\newcommand{\rb}{\right\}}
\newcommand{\lf}{\left.}
\newcommand{\ri}{\right.}
\newcommand{\R}{\mathbb{R}}

\newcommand{\bx}{\mathbf{x}}

\newcommand{\by}{\mathbf{y}}
\newcommand{\bX}{\mathbf{X}}

\newcommand{\bmu}{\boldsymbol\mu}
\newcommand{\bu}{\mathbf{u}}

\newcommand{\bb}{\mathbf{b}}

\newcommand{\bz}{\mathbf{z}}

\newcommand{\bxi}{\boldsymbol\xi}

\newcommand{\ba}{\boldsymbol\alpha}

\newcommand{\bga}{\boldsymbol\gamma}

\newcommand{\Hcal}{\mathcal{H}}
\newcommand{\bHcal}{\boldsymbol{\mathcal{H}}}

\newcommand{\bnab}{\boldsymbol\nabla}
\newcommand{\bB}{\mathbf{B}}

\newcommand{\bI}{\mathbf{I}}

\newcommand{\lrp}[1]{\left(#1\right)}
\newcommand{\lrc}[1]{\left[#1\right]}
\newcommand{\lrb}[1]{\left\{#1\right\}}

\newcommand{\E}[1]{\mathbb{E}\lc #1\rc}
\newcommand{\V}[1]{\mathbb{V}\mathrm{ar}\lc #1\rc}

\newcommand{\mse}[1]{\mathrm{MSE}\lrc{#1}}
\newcommand{\mise}[1]{\mathrm{MISE}\lrc{#1}}
\newcommand{\amise}[1]{\mathrm{AMISE}\lrc{#1}}
\newcommand{\pf}[2]{\frac{\partial #1}{\partial #2}}
\newcommand{\pftwo}[2]{\frac{\partial^2 #1}{\partial #2^2}}
\newcommand{\pfmix}[3]{\frac{\partial^2 #1}{\partial #2\partial #3}}
\newcommand{\norm}[1]{\left|\left| #1\right|\right|}
\newcommand{\abs}[1]{\left| #1\right|}
\newcommand{\tr}[1]{\mathrm{tr}\left[#1\right]}

\newcommand{\vlinel}[1]{\multicolumn{1}{|c}{#1}}

\newcommand{\Om}[1]{\Omega_{#1}}
\newcommand{\om}[1]{\omega_{#1}}
\newcommand{\Iq}[2]{\int_{\Omega_{q}} #1\,\omega_{q}(d #2)}
\newcommand{\Iqr}[3]{\int_{\Omega_{q}\times\R} #1\,d #3\,\omega_{q}(d #2)}

\newcommand{\Ir}[2]{\int_{\R} #1\,d #2}

\DeclareFontFamily{OT1}{pzc}{}
\DeclareFontShape{OT1}{pzc}{m}{it}{<-> s * [1.10] pzcmi7t}{}
\DeclareMathAlphabet{\mathpzc}{OT1}{pzc}{m}{it}
\newcommand{\order}[1]{\mathpzc{o}\lp#1\rp}
\newcommand{\Order}[1]{\mathcal{O}\lp#1\rp}

\usepackage[pdftex,final,bookmarksnumbered,bookmarksopen=false,breaklinks,colorlinks]{hyperref}
\hypersetup{
	final,
    bookmarksnumbered=true,
    bookmarksopen=true,
    bookmarksopenlevel=0,
    unicode=false,          
    pdftoolbar=true,        
    pdfmenubar=true,        
    pdffitwindow=false,     
    pdftitle={Kernel density estimation for directional-linear data},    
	pdfdisplaydoctitle=true, 
	pdftoolbar=true,
	pdfmenubar=true,
	pdflang={English},
    pdfauthor={Eduardo Garcia-Portugues, Rosa M. Crujeiras, Wenceslao Gonzalez-Manteiga},     
    pdfsubject={arXiv paper},   
    pdfcreator={Eduardo Garcia-Portugues},   
    pdfproducer={Eduardo Garcia-Portugues}, 
    pdfkeywords={Directional-linear data}{Kernel density estimator}{Nonparametric statistics}, 
    pdfnewwindow=true,      
    breaklinks=true,
    hidelinks,       
    linkcolor=black,          
    citecolor=black,        
    filecolor=magenta,      
    urlcolor=cyan           
}

\newcommand{\co}{\addtocounter{equation}{1}\arabic{equation}}

\newtheorem{theo}{Theorem}
\newtheorem{coro}{Corollary}
\newtheorem{rem}{Remark}
\newtheorem{prop}{Proposition}
\newtheorem{lem}{Lemma}

\allowdisplaybreaks

\begin{document}

\title{Kernel density estimation for directional-linear data}
\setlength{\droptitle}{-1cm}
\predate{}%
\postdate{}%
\author{Eduardo Garc\'ia-Portugu\'es$^{1,2}$, Rosa M. Crujeiras$^1$, and Wenceslao Gonz\'alez-Manteiga$^1$}

\date{}

\footnotetext[1]{
Department of Statistics and Operations Research, University of Santiago de Compostela (Spain).}
\footnotetext[2]{Corresponding author. e-mail: \href{mailto:eduardo.garcia@usc.es}{eduardo.garcia@usc.es}.}

\maketitle


\begin{abstract}
A nonparametric kernel density estimator for directional-linear data is introduced. The proposal is based on a product kernel accounting for the different nature of both (directional and linear) components of the random vector. Expressions for bias, variance and Mean Integrated Squared Error (MISE) are derived, jointly with an asymptotic normality result for the proposed estimator. For some particular distributions, an explicit formula for the MISE is obtained and compared with its asymptotic version, both for directional and directional-linear kernel density estimators. In this same setting a closed expression for the bootstrap MISE is also derived.
\end{abstract}
\begin{flushleft}
\end{flushleft}

\section{Introduction}
\label{kerdirlin:sec:intro}

Kernel density estimation, and kernel smoothing methods in general, is a classical topic in nonparametric statistics. Starting from the first papers by \cite{Akaike1954}, \cite{Rosenblatt1956} and \cite{Parzen1962}, extensions of the kernel density methodology have been brought up in different contexts, dealing with other smoothers, more complex data (censorship, truncation, dependence) or dynamical models (see \cite{Mueller2006} for a review). Some comprehensive references in this topic include the books by \cite{Silverman1986}, \cite{Scott1992} and \cite{Wand1995}, among others.\\

Beyond the linear case, kernel density estimation has been also adapted to directional data, that is, data in the $q$-dimensional sphere (see \cite{Jupp1989} for a complete review of the theory of directional statistics). \cite{Hall1987} defined two type of kernel estimators and give asymptotic formulae of bias, variance and square loss. Almost simultaneously, \cite{Bai1988} established the pointwise, uniformly strong consistency and $\mathcal{L}_1$ consistency of a quite similar estimator in the same context. Later, \cite{Zhao2001} stated a central limit theorem for the integrated squared error of the previous kernel density estimator based on the $U$-statistic martingale ideas developed by \cite{Hall1984}. Some of the results by \cite{Hall1987} were extended by \cite{Klemela2000}, who studied the estimation of the Laplacian of the density and other types of derivatives. All these references consider the data lying on a general $q$-sphere of arbitrary dimension $q$, which comprises as particular cases circular data ($q=1$) and spherical data ($q=2$). For the particular case of circular data, there are more recent works dealing with the problem of smoothing parameter selection in kernel density estimation, such as \cite{Taylor2008} and \cite{Oliveira2012}. \cite{DiMarzio2011} study the kernel density estimator on the $q$-dimensional torus, and propose some bandwidth selection methods. A more general approach has been followed by \cite{Hendriks1990}, who discusses the estimation of the underlying distribution by means of Fourier expansions in a Riemannian manifold. This differential geometry viewpoint has been exploited recently by \cite{Pelletier2005} and \cite{Henry2009}. Nevertheless, the original approach seems to present a good balance between generality and complexity.\\

The aim of this work is to introduce and derive some basic properties of a joint kernel density estimator for directional-linear data, \textit{i.e.} data with a directional and a linear component. This type of data arise in a variety of applied fields such as meteorology (when analysing the relation between wind direction and wind speed), oceanography (in the study of sea currents) and environmental sciences, among others. As an example, such an estimator has been used by \cite{Garcia-Portugues:so2} for studying the relation between pollutants and wind direction in the presence of an emission source. Specifically, the novelty of this work comprises the analysis of asymptotic properties of the directional-linear kernel density estimator, deriving bias, variance and asymptotic normality. As a by-product, the Mean Integrated Squared Error (MISE) follows, as well as the expression for optimal Asymptotic MISE (AMISE) bandwidths. In addition, for a particular class of densities consisting of mixtures of directional von Mises and normals, it is possible to compare the AMISE with the exact MISE. These results have been also obtained for the purely directional case, considering mixtures of von Mises distributions in the $q$-dimensional sphere, completing the existing results for directional\nolinebreak[4] data. \\

This paper is organized as follows. Section \ref{kerdirlin:sec:background} presents some background on kernel density estimation for linear data and directional data. The proposed directional-linear kernel density estimator and the main results of this paper are included in Section \ref{kerdirlin:sec:mainresults}, where the bias, variance and asymptotic normality are derived. Section \ref{kerdirlin:sec:error} is focused in the issue of error measurement and expressions for the AMISE of the estimator and the exact MISE for particular cases of mixtures are obtained, both in the directional and directional-linear contexts. Conclusions and final comments are given in Section \ref{kerdirlin:sec:conclusions}. The proofs of the results and some technical lemmas are given in the Appendix.

\section[Background on linear and directional kernel density estimation]{Background on linear and directional kernel density\\ estimation}
\label{kerdirlin:sec:background}

This section is devoted to a brief introduction on kernel density estimation for linear and directional data. For the sake of simplicity, $f$ will denote the target density in this paper, which may be linear, directional, or directional-linear, depending on the context.\\

Let $Z$ denote a linear random variable with support $\mathrm{supp}(Z)\subseteq\mathbb R$ and density $f$. Consider $Z_1,\ldots,Z_n$ a random sample of $Z$, with size $n$. The linear kernel density estimator introduced by \cite{Akaike1954}, \cite{Rosenblatt1956} and \cite{Parzen1962} is defined as
\begin{align}
\hat f_g(z)=\frac{1}{ng}\sum_{i=1}^n K\lrp{\frac{z-Z_i}{g}},\quad z\in \R,
\label{kerdirlin:kernel_linear}
\end{align}
where $K$ denotes the kernel, usually a symmetric density about the origin, and $g>0$ is the bandwidth parameter, which controls the smoothness of the estimator. Specifically, large values of the bandwidth parameter will produce oversmoothed estimates of $f$, whereas small values will provide undersmoothed curves. The asymptotic properties of this estimator and its adaptation to different contexts yielded a remarkably prolific field within the statistical literature, as noted in the introduction. \\

It is well known that under some regularity conditions on the kernel and the target density, the bias of the estimator (\ref{kerdirlin:kernel_linear}) is of order $\mathcal O(g^2)$, whereas the variance is $\mathcal O((ng)^{-1})$, clearly showing the need of accounting for a trade-off between bias and variance in any bandwidth selection procedure. Specifically, the expected value of the linear kernel estimator at $z\in\mathbb R$ is:
\begin{align*}
\E{\hat f_g(z)}=f(z)+\frac{1}{2}\mu_2(K)f''(z)g^2+\order{g^2},
\end{align*}
where $\mu_p(K)=\int_\R z^p K(z)\,dz$ represents the $p$-th moment of the kernel $K$. Similarly, the variance of (\ref{kerdirlin:kernel_linear}) at $z\in\mathbb R$ is given by:
\begin{align*}
\V{\hat f_g(z)}=(ng)^{-1}R(K)f(z)+\order{(ng)^{-1}},
\end{align*}
where $R(K)=\int_\R K^2(z)\,dz$. Further details on computations for the linear kernel density estimator can be found in Section 2.5 of \cite{Wand1995}.

\subsection{Kernel density estimation for directional data}
\label{kerdirlin:subsec:kdedir}

As previously mentioned, kernel density estimation has been adapted to different contexts such as directional data, that is, data on a $q$-dimensional sphere, being circular data ($q=1$) and spherical data ($q=2$) particular cases. Let $\bX$ denote a directional random variable with density $f$. The support of such a variable is the $q$-dimensional sphere, denoted by $\Om{q}=\big\{\bx\in\R^{q+1}:x^2_1+\cdots+x^2_{q+1}=1\big\}$. The Lebesgue measure in $\Om{q}$ will be denoted by $\om{q}$ and, therefore, a directional density satisfies
\[
\Iq{f(\bx)}{\bx}=1.
\]
\begin{rem}
When there is no possible misunderstanding, $\om{q}$ will also denote the surface area \nolinebreak[4]of\nolinebreak[4] $\Om{q}$:
\[
\om{q}=\om{q}\lrp{\Om{q}}=\frac{2\pi^\frac{q+1}{2}}{\Gamma\lrp{\frac{q+1}{2}}},\quad q\geq 1, 
\]
where $\Gamma$ represents the Gamma function defined as $\Gamma(p)=\int_0^\infty x^{p-1}e^{-x}\,dx$, for $p>-1$. 
\end{rem}

The directional kernel density estimator was proposed by \cite{Hall1987} and \cite{Bai1988}, following two different perspectives in the treatment of directional data. In this paper, the definition in \cite{Bai1988} will be considered, although it can also be related with one of the proposals in \cite{Hall1987}. Given a random sample $\bX_1,\ldots,\bX_n$, of a directional variable $\bX$ with density $f$, the directional kernel density estimator is given by:
\begin{align}
\hat f_h(\bx)=\frac{c_{h,q}(L)}{n}\sum_{i=1}^n L\lrp{\frac{1-\bx^T\bX_i}{h^2}},\quad \bx\in \Omega_q,
\label{kerdirlin:kernel_directional}
\end{align}
where $L$ is the directional kernel, $h>0$ is the bandwidth parameter and $c_{h,q}(L)$ is a normalizing constant depending on the kernel $L$, the bandwidth $h$ and the dimension $q$. The scalar product of two vectors, $\bx$ and $\by$, is denoted by $\bx^T\by$, where $^T$ is the transpose operator.\\

In this setting, directional kernels are not directional densities but functions of rapid decay. Therefore, to ensure that the resulting estimator is indeed a directional density, the normalizing constant $c_{h,q}(L)$ is needed. Specifically (see \cite{Bai1988}), the inverse of this normalizing constant for any $\bx\in\Omega_q$ is given by
\begin{align}
c_{h,q}(L)^{-1}=\Iq{L\lrp{\frac{1-\bx^T\by}{h^2}}}{\by}=h^q\lambda_{h,q}(L)\sim h^q\lambda_{q}(L),
\label{kerdirlin:normalizing}
\end{align}
with $\lambda_{h,q}(L)=\om{q-1}\int_0^{2h^{-2}} L(r) r^{\frac{q}{2}-1}(2-rh^2)^{\frac{q}{2}-1}\,dr$ and $\lambda_q(L)=2^{\frac{q}{2}-1}\om{q-1}\int_0^{\infty} L(r) r^{\frac{q}{2}-1}\,dr$.  The asymptotic behaviour of $\lambda_{h,q}(L)$ is established in Lemma \ref{kerdirlin:dir:lem:1a} and the notation $a_n\sim b_n$ indicates that $\frac{a_n}{b_n}\rightarrow 1$ as $n\to\infty$ (see also \cite{Bai1988} and \cite{Zhao2001}). \\

Properties of the directional kernel density estimator (\ref{kerdirlin:kernel_directional}) have been analysed by \cite{Bai1988}, who proved pointwise, uniform and $\mathcal{L}_1$-norm consistency. A central limit theorem for the integrated squared error of the estimator has been established by \cite{Zhao2001}, as well as the expression for the bias under some regularity conditions, stated below:

\begin{enumerate}[label=\textbf{D\arabic{*}}.,ref=\textbf{D\arabic{*}}]

\item Extend $f$ from $\Omega_q$ to $\R^{q+1}\backslash\lrb{\mathbf{0}}$ by defining $f(\bx)\equiv f\lrp{\bx/\norm{\bx}}$ for all $\bx\neq\mathbf{0}$, where $\norm{\cdot}$ denotes the Euclidean norm. Assume that the gradient vector $\bnab f(\bx)=\lrp{\pf{f(\bx)}{x_1},\cdots,\pf{f(\bx)}{x_{q+1}}}^T$ and the Hessian matrix $\bHcal f(\bx)=\lrp{\pfmix{f(\bx)}{x_i}{x_j}}_{1\leq i,j\leq q+1}$ exist and are continuous on $\R^{q+1}\backslash\lrb{\mathbf{0}}$. \label{kerdirlin:cond:d1}
 
\item Assume that $L:[0,\infty)\rightarrow[0,\infty)$ is a bounded and Riemann integrable function such that\label{kerdirlin:cond:d2}
\[
0<\int_0^\infty L^k(r) r^{\frac{q}{2}-1}\,dr<\infty,\quad\forall q\geq1,\mbox{ for } k=1,2.
\]
\item Assume that $h=h_n$ is a sequence of positive numbers such that $h_n\rightarrow0$ and $n h_n^q\rightarrow\infty$ as $n\rightarrow\infty$.\label{kerdirlin:cond:d3}
\end{enumerate}

\begin{rem}
\label{kerdirlin:dir:rem:1}
$L$ must be a rapidly decreasing function, quite different from the bell-shaped kernels $K$ involved in the linear estimator (\ref{kerdirlin:kernel_linear}). To verify \ref{kerdirlin:cond:d2}, $L$ must decrease faster than any power function, since $\int_0^\infty r^\alpha r^{\frac{q}{2}-1}\,dr=\infty$, $\forall \alpha\in\R$, $\forall q\geq1$.
\end{rem}

Lemma 2 in \cite{Zhao2001} states that, under the previous conditions \ref{kerdirlin:cond:d1}--\ref{kerdirlin:cond:d3}, the expected value of the directional kernel density estimator in a point $\bx\in\Om{q}$, is
\begin{align*}
\E{\hat f_h(\bx)}=f(\bx)+b_q(L)\Psi(f,\bx)h^2+\order{h^2},
\end{align*}
where
\begin{align}
\Psi(f,\bx)=&\,-\bx^T\bnab f(\bx)+q^{-1}\lrp{\nabla^2f(\bx)-\bx^T\bHcal f(\bx)\bx},\label{kerdirlin:Psi_dir}\\
\quad b_q(L)=&\,\int_0^\infty L(r) r^{\frac{q}{2}}\,dr\bigg/\int_0^\infty L(r) r^{\frac{q}{2}-1}\,dr\label{kerdirlin:bq},
\end{align}
being $\nabla^2 f(\bx)=\sum_{i=1}^{q+1} \frac{\partial^2 f(\bx)}{\partial \bx_i^2}$ the Laplacian of $f$. Note that the bias is of order $\mathcal O(h^2)$, but in (\ref{kerdirlin:Psi_dir}), apart from the curvature of the target density which is captured by the Hessian matrix, a gradient vector also appears. On the other hand, the scaling constant $b_q(L)$ can be interpreted as a kind of moment of the directional kernel $L$. Note that, condition \ref{kerdirlin:cond:d2} with $k=1$ is needed for the bias computation. The same condition with $k=2$ is required for deriving the pointwise variance of the estimator (\ref{kerdirlin:kernel_directional}), which was also given by \cite{Hall1987} and \cite{Klemela2000}.

\begin{prop}
\label{kerdirlin:dir:prop:2}
Under conditions \ref{kerdirlin:cond:d1}--\ref{kerdirlin:cond:d3}, the variance of $\hat f_h(\bx)$ at $\bx\in\Omega_q$ is given by
\begin{align*}
\V{\hat f_h(\bx)}=\frac{c_{h,q}(L)}{n}d_q(L)f(\bx)+\order{(nh^q)^{-1}},
\end{align*}
where
\begin{align*}
d_q(L)=\int_0^\infty L^2(r) r^{\frac{q}{2}-1}\,dr\bigg/\int_0^\infty L(r) r^{\frac{q}{2}-1}\,dr.
\end{align*}
\end{prop}

Regarding the normalizing constant expression (\ref{kerdirlin:normalizing}), the order of the variance is $\Order{(nh^q)^{-1}}$, where $q$ is the dimension of the sphere. This order coincides with the corresponding one for a multivariate kernel density estimator in $\R^q$ (see \cite{Scott1992}).\\ 

\begin{figure}[h]
	\vspace{-0.25cm}
	\centering
	\includegraphics[width=0.45\textwidth]{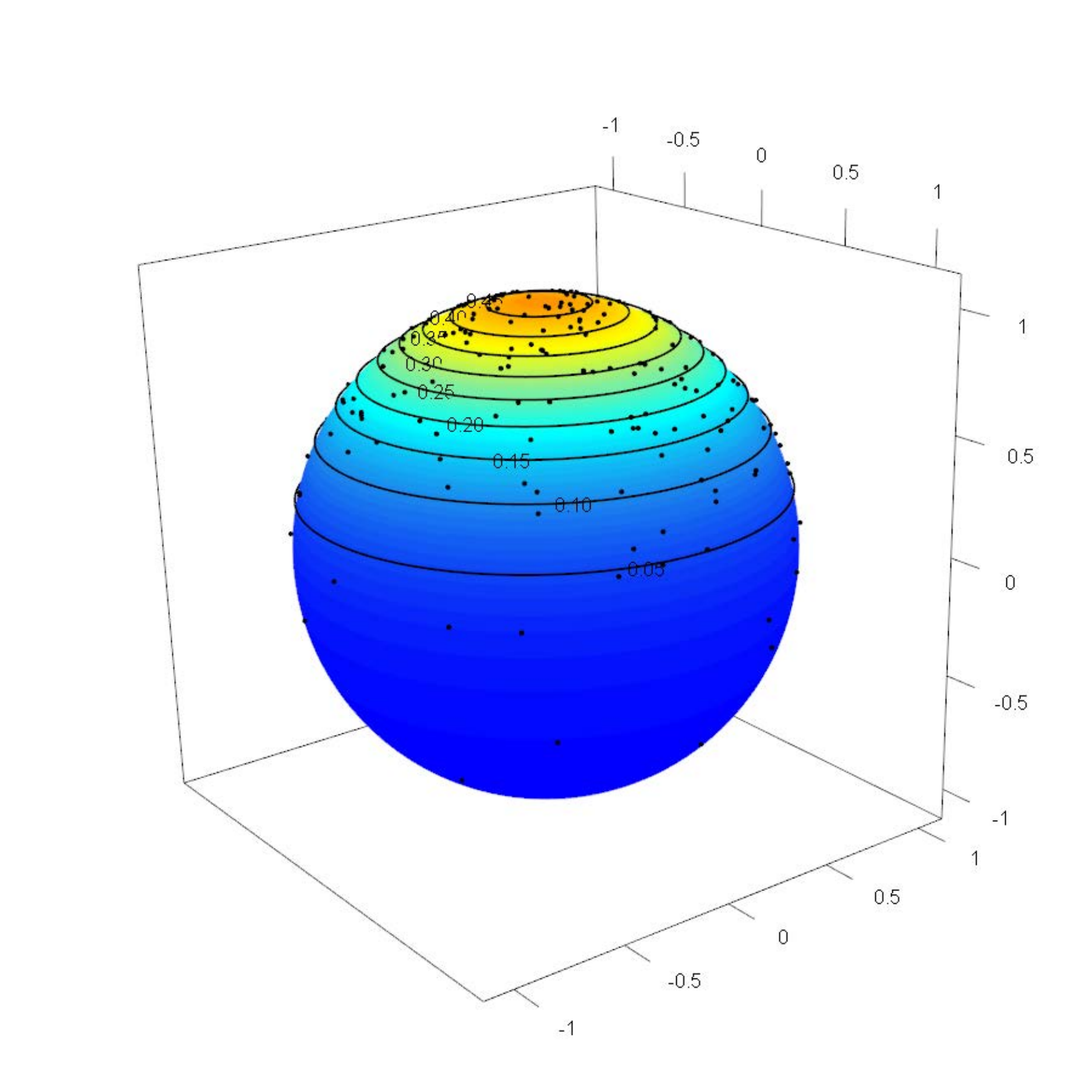}
	\includegraphics[width=0.45\textwidth]{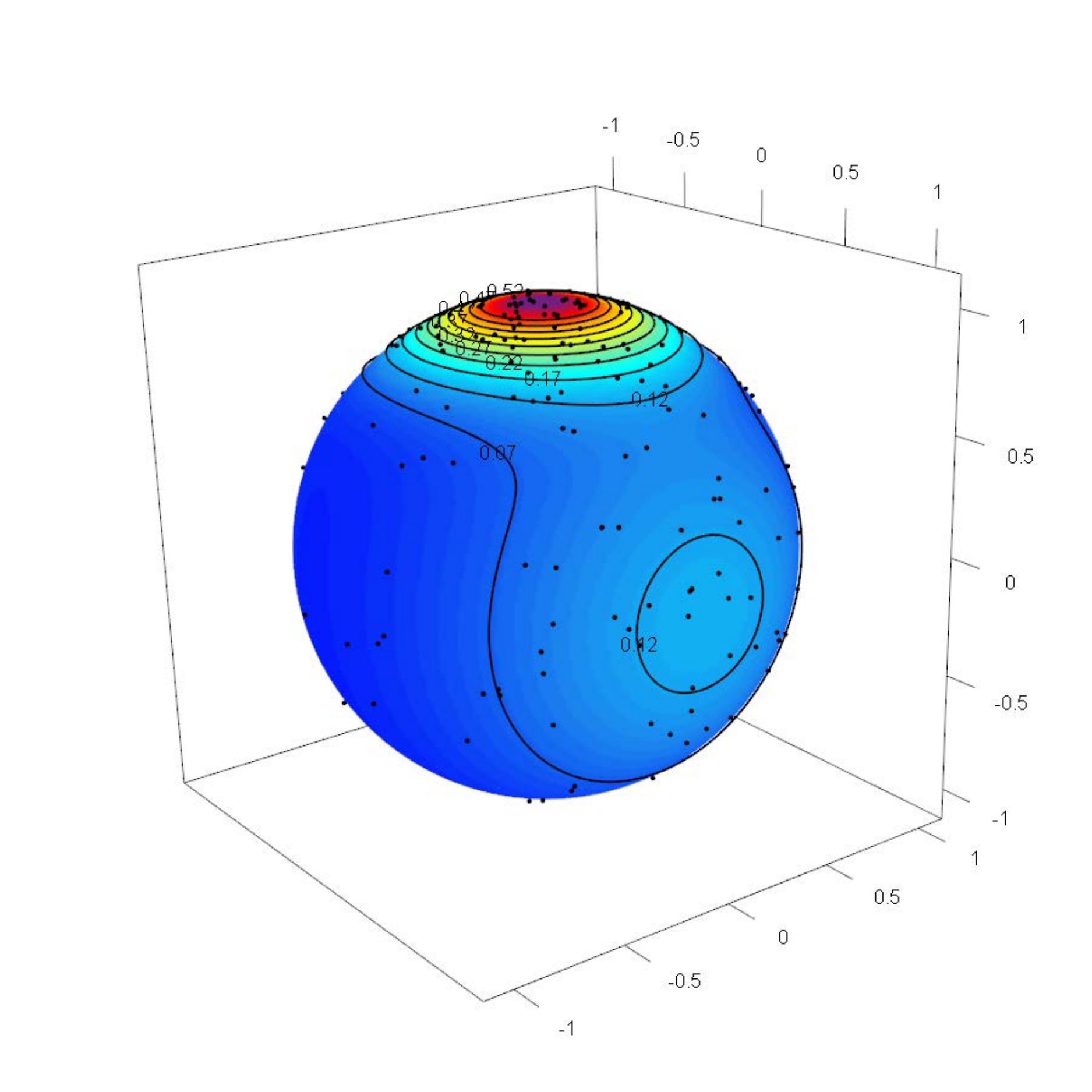}
	\caption{\small Left: contour plot of a von Mises density $\mathrm{vM}(\bmu,\kappa)$, with $\bmu=(0,0,1)$ and $\kappa=3$. Right: contour plot of the mixture of von Mises densities (\ref{kerdirlin:mixtdir}). \label{kerdirlin:fig0}}
\end{figure}

A popular choice for the directional kernel is $L(r)=e^{-r}$, $r\geq0$, also known as the von Mises kernel due to its relation with the von Mises--Fisher distribution (see \cite{Watson1983}). In a $q$-dimensional sphere, the von Mises model $\mathrm{vM}(\bmu,\kappa)$ has density
\begin{align}
f_{\mathrm{vM}}(\bx;\bmu,\kappa)=C_q(\kappa) \exp{\lrb{\kappa\bx^T\bmu}},\quad C_q(\kappa)=\frac{\kappa^{\frac{q-1}{2}}}{(2\pi)^{\frac{q+1}{2}}\mathcal{I}_{\frac{q-1}{2}}(\kappa)},\label{kerdirlin:dir:vm}
\end{align}
being $\bmu\in\Omega_q$ the directional mean and $\kappa\geq0$ the concentration parameter around the mean. In Figure \ref{kerdirlin:fig0} (left plot), the contour plot of a spherical von Mises is shown. $\mathcal{I}_\nu$ is the modified Bessel function of order $\nu$,
\begin{align*}
\mathcal{I}_\nu(z)=\frac{\lrp{\frac{z}{2}}^\nu}{\pi^{1/2}\Gamma\lrp{\nu+\frac{1}{2}}}\int_{-1}^1 (1-t^2)^{\nu-\frac{1}{2}}e^{zt}\,dt.
\end{align*}

For the particular case of the target density being a $q$-dimensional von Mises $\mathrm{vM}(\bmu,\kappa)$, the term (\ref{kerdirlin:Psi_dir}) in the bias computation becomes:
\begin{align*}
\Psi\lrp{f_{\mathrm{vM}}(\cdot;\bmu,\kappa),\bx}=\kappa C_q(\kappa)e^{\kappa\bx^T\bmu}\lrp{-\bx^T\bmu+\kappa q^{-1}\lrp{1-(\bx^T\bmu)^2}}.
\end{align*}
As $\kappa\rightarrow 0$, which means that the distribution is approaching a uniform model in the sphere, the previous term also tends to zero.\\ 

Considering the von Mises kernel in the directional estimator (\ref{kerdirlin:kernel_directional}) allows for its interpretation as a mixture of von Mises--Fisher densities
\begin{align}
\hat f_h(\bx)=\frac{1}{n}\sum_{i=1}^n f_{\mathrm{vM}}\lrp{\bx;\bX_i,1/h^2},\label{kerdirlin:kernel_vonmises}
\end{align}
where, for each von Mises component, the mean value is $i$-th observation $\bX_i$ and the concentration is given by $\frac{1}{h^2}$, involving the smoothing parameter.

In addition, the normalizing constant (\ref{kerdirlin:normalizing}) appearing in the construction of the directional kernel estimator (\ref{kerdirlin:kernel_directional}) has a simple expression for a von Mises kernel, given by
\begin{align}
c_{h,q}(L)^{-1}=\frac{2\pi^{\frac{q}{2}}}{\Gamma\lrp{\frac{q}{2}}} \int_{-1}^1 \exp\lrb{\frac{-1+t}{h^2}}(1-t^2)^{\frac{q}{2}-1}\,dt=C_q(1/h^2)^{-1}e^{-1/h^2}.
\label{kerdirlin:norm_von_Mises}
\end{align}

For a general kernel, the asymptotic behaviour of $c_{h,q}(L)^{-1}$ was remarked in (\ref{kerdirlin:normalizing}) and it can be specified for the von Mises kernel. In this case, (\ref{kerdirlin:norm_von_Mises}) depends on $C_q(1/h^2)$, which involves a Bessel function of order $(q-1)/2$. Applying a Taylor expansion for $\mathcal{I}_\nu$, it can be seen that $\mathcal{I}_\nu(z)=e^z\Big(\frac{z^{-\frac{1}{2}}}{\sqrt{2\pi}}+\mathcal{O}\big(z^{-\frac{3}{2}}\big)\Big)$, $z\geq0$ and $c_{h,q}(L)^{-1}$ presents also a simple form:
\begin{align*}
c_{h,q}(L)^{-1}=\lrp{2\pi}^{\frac{q+1}{2}}e^{-\frac{1}{h^2}}h^{q-1}e^\frac{1}{h^2}\lrp{\frac{h}{\sqrt{2\pi}}+\Order{h^3}}=\lrp{2\pi}^{\frac{q}{2}}h^q+\Order{h^{q+2}}.
\end{align*}

Finally, the other terms involved in bias and variance, namely $b_q(L)$ and $d_q(L)$, become
\begin{align*}
b_q(L)=\frac{q}{2},\quad d_q(L)=2^{-\frac{q}{2}},\quad\forall q\geq 1
\end{align*}
for the von Mises kernel.

\subsection{Kernel density estimation for directional-linear data}
\label{kerdirlin:subsec:kdedirlin}

Consider a directional-linear random variable, $(\bX,Z)$ with support $\mathrm{supp}(\bX,Z)\subseteq \Omega_q\times\mathbb R $ and joint density $f$. For the simple case of circular data ($q=1$), the support of the variable is the cylinder. Following the ideas in the previous section for the linear and directional cases, given a random sample $\lrp{\bX_1,Z_1},\ldots,\lrp{\bX_n,Z_n}$, the directional-linear kernel density estimator can be defined as:
\begin{align}
\hat f_{h,g}(\bx,z)=\frac{c_{h,q}(L)}{ng}\sum_{i=1}^nLK\lp\frac{1-\bx^T\bX_i}{h^2},\frac{z-Z_i}{g}\rp,\quad (\bx,z)\in\Omega_q\times\mathbb R,
\label{kerdirlin:kernel_dirlinear}
\end{align}
where $LK$ is a directional-linear kernel, $g$ is the bandwidth parameter for the linear component, $h$ the bandwidth parameter for the directional component and $c_{h,q}(L)$ is the normalizing constant for the directional part, defined in (\ref{kerdirlin:normalizing}). For the sake of simplicity, a product kernel $LK(\cdot,\cdot)=L(\cdot)\times K(\cdot)$ will be considered throughout this paper. Although a product kernel formulation has been adopted, the results could be generalized for a directional-linear kernel, with the suitable modifications in the required conditions. 

\section{Main results}
\label{kerdirlin:sec:mainresults}

Before stating the main results, some notation will be introduced. The target directional-linear density will be denoted by $f$. The gradient vector and Hessian matrix of $f$, with respect to both components (directional and linear) are defined in this setting as:
\begin{align*}
\bnab f(\bx,z)=&\,\lrp{\pf{f(\bx,z)}{x_1},\ldots,\pf{f(\bx,z)}{x_{q+1}},\pf{f(\bx,z)}{z}}^T=\lrp{\bnab_{\bx}f(\bx,z),\nabla_z f(\bx,z)}^T,\\
\renewcommand{\arraystretch}{1.4} %
\bHcal f(\bx,z)=&\,
\lrp{\begin{array}{cccc}
\pftwo{f(\bx,z)}{x_1} & \cdots & \pfmix{f(\bx,z)}{x_1}{x_{q+1}} & \vlinel{\pfmix{f(\bx,z)}{x_1}{z}} \\
\vdots & \ddots & \vdots & \vlinel{\vdots} \\
\pfmix{f(\bx,z)}{x_{q+1}}{x_1} & \cdots & \pftwo{f(\bx,z)}{x_{q+1}} & \vlinel{\pfmix{f(\bx,z)}{x_{q+1}}{z}} \\[0.1cm]\cline{1-4}
& & & \vlinel{}\\[-0.35cm]
\pfmix{f(\bx,z)}{z}{x_1} & \cdots & \pfmix{f(\bx,z)}{z}{x_{q+1}} & \vlinel{\pftwo{f(\bx,z)}{z}}
\end{array}}=
\lrp{\begin{array}{cc}
\bHcal_\bx f(\bx,z) & \vlinel{\bHcal_{\bx,z} f(\bx,z)} \\[0.1cm]\cline{1-2}
& \vlinel{}\\[-0.35cm]
\bHcal_{\bx,z} f(\bx,z)^T & \vlinel{\Hcal_z f(\bx,z)}
\end{array}},
\end{align*}
where subscripts $\bx$ and $z$ are used to denote the derivatives with respect to the directional and linear components, respectively. The Laplacian of $f$ restricted to the directional component is denoted by $\nabla_\bx^2 f(\bx,z)=\sum_{i=1}^{q+1} \frac{\partial^2 f(\bx,z)}{\partial \bx_i^2}$. The following conditions will be required in order to prove the main results:

\begin{enumerate}[label=\textbf{DL\arabic{*}}.,ref=\textbf{DL\arabic{*}}]
\item Extend $f$ from $\Omega_q\times\R$ to $\R^{q+2}\backslash A$, $A=\lrb{(\bx,z)\in\R^{q+2}:\bx=\mathbf{0}}$, by defining $f(\bx,z)\equiv f\lrp{\bx/\norm{\bx},z}$ for all $\bx\neq\mathbf{0}$ and $z\in\R$, where $\norm{\cdot}$ denotes the Euclidean norm. Assume that $\bnab f(\bx,z)$ and $\bHcal f(\bx,z)$ exist, are continuous and square integrable on $\Om{q}\times\mathbb{R}$. \label{kerdirlin:cond:dl1}
 
\item Assume that the directional kernel $L$ satisfies condition \ref{kerdirlin:cond:d2} and the linear kernel $K$ is a symmetric around zero and bounded linear density function with finite second order moment.\label{kerdirlin:cond:dl2}

\item  Assume that $h=h_n$ and $g=g_n$ are sequences of positive numbers such that $h_n\rightarrow0$, $g_n\rightarrow0$ and $n h_n^q g_n\rightarrow\infty$ as $n\rightarrow\infty$.\label{kerdirlin:cond:dl3}

\end{enumerate}

The next two results provide the expressions for the bias and the variance of the directional-linear kernel density estimator (\ref{kerdirlin:kernel_dirlinear}). 

\begin{prop}
\label{kerdirlin:dirlin:prop:1}
Under conditions \ref{kerdirlin:cond:dl1}--\ref{kerdirlin:cond:dl3}, the expected value of the directional-linear kernel density estimator (\ref{kerdirlin:kernel_dirlinear}) in a point $(\bx,z)\in\Omega_{q}\times\R$ is given by
\begin{align*}
\E{\hat f_{h,g}(\bx,z)}=f(\bx,z)+b_q(L)\Psi_\bx(f,\bx,z)h^2+\frac{1}{2}\mu_2(K)\Hcal_z f(\bx,z)g^2+\order{h^2+g^2},
\end{align*}
where
\begin{align*}
\Psi_\bx(f,\bx,z)=-\bx^T\bnab_\bx f(\bx,z)+q^{-1}\lrp{\nabla_\bx^2f(\bx,z)-\bx^T\bHcal _\bx f(\bx,z)\bx}.
\end{align*}
\end{prop}

\begin{prop}
\label{kerdirlin:dirlin:prop:2}
Under conditions \ref{kerdirlin:cond:dl1}--\ref{kerdirlin:cond:dl3}, the variance for the directional-linear kernel density estimator (\ref{kerdirlin:kernel_dirlinear}) in a point $(\bx,z)\in\Omega_{q}\times\R$ is given by
\begin{align*}
\V{\hat f_{h,g}(\bx,z)}=\frac{c_{h,q}(L)}{ng}R(K)d_q(L)f(\bx,z)+\order{(nh^qg)^{-1}}.
\end{align*}
\end{prop}

In view of the previous results, some comments must be done. Firstly, the effects of the directional and linear part can be clearly identified. For the bias, marginal contributions appear as two addends and also the remaining orders from each part are separated. For the variance, the terms corresponding to both parts can be also identified, although turning up in a product form. In addition, the respective orders for bias and variance are analogous to those ones obtained with a $(q+1)$-multivariate estimator in $\R^{q+1}$ (see \cite{Scott1992}).\\

It can be also proved that the directional-linear kernel density estimator (\ref{kerdirlin:kernel_dirlinear}) is asymptotically normal, under the same conditions as those ones used for deriving the expected value and the variance, and a further smoothness property on the product kernel.

\begin{theo}
\label{kerdirlin:normality:th:1}
Under conditions \ref{kerdirlin:cond:dl1}--\ref{kerdirlin:cond:dl3}, if $\displaystyle\int_0^{\infty}\int_\R LK^{2+\delta}\lrp{r,v}r^{\frac{q}{2}-1}\,dv\,dr<\infty$ for some $\delta>0$, then the directional-linear kernel density estimator (\ref{kerdirlin:kernel_dirlinear}) is asymptotically normal:
\begin{align*}
\sqrt{nh^qg}\lrp{\hat f_{h,g}(\bx,z)-f(\bx,z)-\mathrm{ABias}\lrc{\hat f_{h,g}(\bx,z)}}\stackrel{d}{\longrightarrow}\mathcal{N}\lrp{0,R(K)d_q(L)f(\bx,z)},
\end{align*}
pointwise in $(\bx,z)\in\Om{q}\times\R$, where $\mathrm{ABias}\big[\hat f_{h,g}(\bx,z)\big]=b_q(L)\Psi_\bx(f,\bx,z)h^2+\frac{1}{2}\mu_2(K)\Hcal_z f(\bx,z)g^2$.
\end{theo}
The smoothness condition on the directional-linear kernel is required in order to ensure Lyapunov's condition and obtain the asymptotic normal distribution. Again, the effect of the two parts can be identified in the previous equation, as well as in the rate of convergence of the estimator.

\section{Error measurement and optimal bandwidth}
\label{kerdirlin:sec:error}

The analysis of the performance of the kernel density estimator requires the specification of appropriate error criteria. Consider a generic kernel density estimator $\hat f$, which can be linear, directional or directional-linear. A global error measurement for quantifying the overall performance of this estimator is given by the MISE:
\begin{align*}
\mathrm{MISE}\lrc{\hat f}=\int\E{(\hat f(u)-f(u))^2}\,du.
\end{align*}
The MISE can be interpreted as a function of the bandwidth and its minimization yields an optimal bandwidth in the sense of the quadratic loss.\\

For the linear kernel density estimator (\ref{kerdirlin:kernel_linear}) and under some regularity conditions (see \cite{Wand1995}), the MISE is given by:
\begin{align*}
\mise{\hat f_g}=&\,\frac{1}{4}\mu_2(K)^2\,R(f'')g^4+(ng)^{-1}R(K)+\order{g^4+(ng)^{-1}}.
\end{align*}
The asymptotic version of the MISE, namely the AMISE, can be used to derive an optimal bandwidth that minimizes this error. This optimal bandwidth is given by
\begin{align*}
g_{\mathrm{AMISE}}=&\,\lrc{\frac{R(K)}{\mu_2(K)^2R(f'')n}}^{\frac{1}{5}}.
\end{align*}
Although the previous expression does not provide a bandwidth value in practice, given that it depends on the curvature of the target density $R(f'')$, some interesting issues should be noticed. For instance, the order of the asymptotic optimal bandwidth is $\mathcal O(n^{-1/5})$. Also, this result is the starting point of more sophisticated bandwidth selectors such as the ones given by \cite{Sheather1991} and \cite{Cao1993}. A comparison of the performance of different bandwidth selectors can be found in \cite{Cao1994}, whereas \cite{Jones1996} provides a review on bandwidth selection methods.

\subsection{\texorpdfstring{MISE for directional and directional-linear kernel density estimators}{MISE for directional and directional-linear kernel density estimators}}

In the previous sections, the bias and variance for the directional kernel estimator (see \cite{Zhao2001} for the bias and Proposition \ref{kerdirlin:dir:prop:2} for the variance) and for the directional-linear kernel estimator (Propositions \ref{kerdirlin:dirlin:prop:1} and \ref{kerdirlin:dirlin:prop:2}) were obtained. Hence, it is straightforward to get the MISE for these estimators. 

\begin{prop}
\label{kerdirlin:dir:prop:3}
Under conditions \ref{kerdirlin:cond:d1}--\ref{kerdirlin:cond:d3}, the MISE for the directional kernel density estimator (\ref{kerdirlin:kernel_directional}) is given by
\begin{align*}
\mise{\hat f_h}=&\,b_q(L)^2\int_{\Omega_{q}}\Psi(f,\bx)^2\,\omega_q(d\bx)h^4+\frac{c_{h,q}(L)}{n}d_q(L)+\order{h^4+(nh^q)^{-1}}.
\end{align*}
\end{prop}

Following \cite{Wand1995}, $\mathrm{MISE}\big[\hat f_h\big]=\mathrm{AMISE}\big[\hat f_h\big]+\order{h^4+(nh^q)^{-1}}$, providing\linebreak  $\mathrm{AMISE}\big[\hat f_h\big]$ a suitable large sample approximation that allows for the computation of an optimal bandwidth with closed expression, minimizing this asymptotic error criterion.

\begin{coro}
\label{kerdirlin:dir:cor:1}
The AMISE optimal bandwidth for the directional kernel density estimator (\ref{kerdirlin:kernel_directional}) is given\nolinebreak[4] by
\begin{align*}
h_{\mathrm{AMISE}}=&\,\lrc{\frac{qd_q(L)}{4b_q(L)^2\lambda_q(L) R(\Psi(f,\cdot)) n}}^{\frac{1}{4+q}},
\end{align*}
where $R(\Psi(f,\cdot))=\Iq{\Psi(f,\bx)^2}{\bx}$ and $\lambda_q(L)=2^{\frac{q}{2}-1}\om{q-1}\int_0^{\infty} L(r) r^{\frac{q}{2}-1}\,dr$. 
\end{coro}

Expressions for MISE and AMISE can be also derived for the directional-linear estimator. In order to simplify the notation, let denote $I\lrc{\phi}=\Iqr{\phi(\bx,z)}{\bx}{z}$, for a function $\phi:\Omega_q\times\R\rightarrow\nolinebreak[4]\R$.  

\begin{prop}
\label{kerdirlin:dirlin:prop:3}
Under conditions \ref{kerdirlin:cond:dl1}--\ref{kerdirlin:cond:dl3}, the MISE for the directional-linear kernel density estimator (\ref{kerdirlin:kernel_dirlinear}) is given by
\begin{align*}
\mise{\hat f_{h,g}}=&\,b_q(L)^2I\lrc{\Psi_\bx(f,\cdot,\cdot)^2} h^4+\frac{1}{4}\mu_2(K)^2I\lrc{\Hcal_z f(\cdot,\cdot)^2}g^4\\
&+b_q(L)\mu_2(K)I\lrc{\Psi_\bx(f,\cdot,\cdot)\Hcal_z f(\cdot,\cdot)}h^2g^2+\frac{c_{h,q}(L)}{ng}d_q(L)R(K)\\
&+\order{h^4+g^4+(nh^qg)^{-1}}.
\end{align*}
\end{prop}

Unfortunately, it is not straightforward to derive a full closed expression for the optimal pair of bandwidths $(h,g)_\mathrm{AMISE}$, although it is possible to compute them by numerical optimization. However, such a closed expression can be obtained for the particular case $q=1$, where the circular and linear bandwidths can be considered as proportional.

\begin{coro}
\label{kerdirlin:dirlin:cor:1}
Consider the parametrization $g=\beta h$. The optimal AMISE pair of bandwidths 
$(h,g)_{\mathrm{AMISE}}=(h_{\mathrm{AMISE}},\beta h_{\mathrm{AMISE}})$ can be obtained from
\begin{align*}
h_{\mathrm{AMISE}}=&\,\lrc{\frac{(q+1)d_q(L)R(K)}{4\beta\lambda_q(L)R\big(b_q(L)\Psi_\bx(f,\cdot,\cdot)+\frac{\beta^2}{2}\mu_2(K)\Hcal_z f(\cdot,\cdot)\big) n}}^{\frac{1}{5+q}},
\end{align*}
where $R\Big(b_q(L)\Psi_\bx(f,\cdot,\cdot)+\frac{\beta^2}{2}\mu_2(K)\Hcal_z f(\cdot,\cdot)\Big)=\int_{\Om{q}\times\R}\big(b_q(L)\Psi_\bx(f,\bx,z)+\frac{\beta^2}{2}\mu_2(K)\Hcal_z f(\bx,z)\big)^2$\\ $\,dz\,\om{q}(d\bx)$ and $\lambda_q(L)$ is defined as in the previous corollary. For the circular-linear data case ($q=1$), the parameter $\beta$ is given by:
\begin{align*}
\beta=\lrp{\frac{\frac{1}{4}\mu_2(K)^2 I\lrc{\Hcal_z f(\cdot,\cdot)^2}}{b_q(L)^2 I\lrc{\Psi_\bx (f,\cdot,\cdot)^2}}}^{\frac{1}{4}}.
\end{align*}
\end{coro}
 
Despite a formal way for deriving the orders of the AMISE bandwidths has not been derived, a quite plausible conjecture is that for $q>1$, $(h,g)_\mathrm{AMISE}=\big(\mathcal{O}\big(n^{-1/(4+q)}\big), \allowbreak\mathcal{O}\big(n^{-1/5}\big)\big)$ or, equivalently, that $\beta=\beta_n=\mathcal{O}\big(n^{-(q-1)/(5(4+q))}\big)$. Indeed, this is satisfied for $q=1$. \\

Finally, it is interesting to note that considering $g=\beta h$, a single bandwidth for the kernel estimator (\ref{kerdirlin:kernel_dirlinear}) is required, having the optimal bandwidth under this formulation order $\mathcal{O}\big(n^{-1/(5+q)}\big)$. This coincides with the order of the kernel linear estimator in $\R^p$, with $p=\dim(\Omega_q\times\R)=q+1$. 

\subsection{\texorpdfstring{Some exact MISE calculations for mixture distributions}{Some exact MISE calculations for mixture distributions}}
\label{kerdirlin:sec:mise}

Closed expressions for the MISE for the directional and directional-linear estimators can be obtained for some particular distribution models, and they will be derived in this section. In the linear setting, \cite{Marron1992} obtained a closed expression for the MISE of (\ref{kerdirlin:kernel_linear}) if the kernel $K$ is a normal density and the underlying model is a mixture of normal distributions. Specifically, the density of an $r$-mixture of normal distributions with respective means $m_j$ and variances $\sigma^2_j$, for $j=1,\ldots,r$ is given by
\begin{align*}
f_r(z)=\sum_{j=1}^r p_j \phi_{\sigma_j}\lrp{z-m_j},\quad \sum_{j=1}^r p_j=1,\quad p_j\geq 0,
\end{align*}
where $p_j$, $j=1,\ldots,r$ denote the mixture weights and $\phi_\sigma$ is the density of a normal with zero mean and variance $\sigma^2$, \textit{i.e.}, $\phi_\sigma(z)=\frac{1}{\sqrt{2\pi}\sigma}\exp\lrb{-\frac{z^2}{2\sigma^2}}$. \cite{Marron1992} showed that the exact MISE of the linear kernel estimator is
\begin{align}
\mise{\hat f_g}=\lrp{2\pi^\frac{1}{2}gn}^{-1}+\mathbf{p}^T\lrc{(1-n^{-1})\mathbf{\Omega_2}(g)-2\mathbf{\Omega_1}(g)+\mathbf{\Omega_0}(g)}\mathbf{p},\label{kerdirlin:mise:no}
\end{align}
where $\mathbf{p}=\lrp{p_1,\ldots,p_r}^T$ and $\mathbf{\Omega_a}(g)$ are matrices with entries $\mathbf{\Omega_a}(g)=\lrp{\phi_{\sigma_a}(m_i-m_j)}_{ij}$, $\sigma_a=\big(ag^2+\sigma_i^2+\sigma_j^2\big)^\frac{1}{2}$, for $a=0,1,2$.\\

\begin{figure}[h]
\centering
\includegraphics[width=0.3\textwidth]{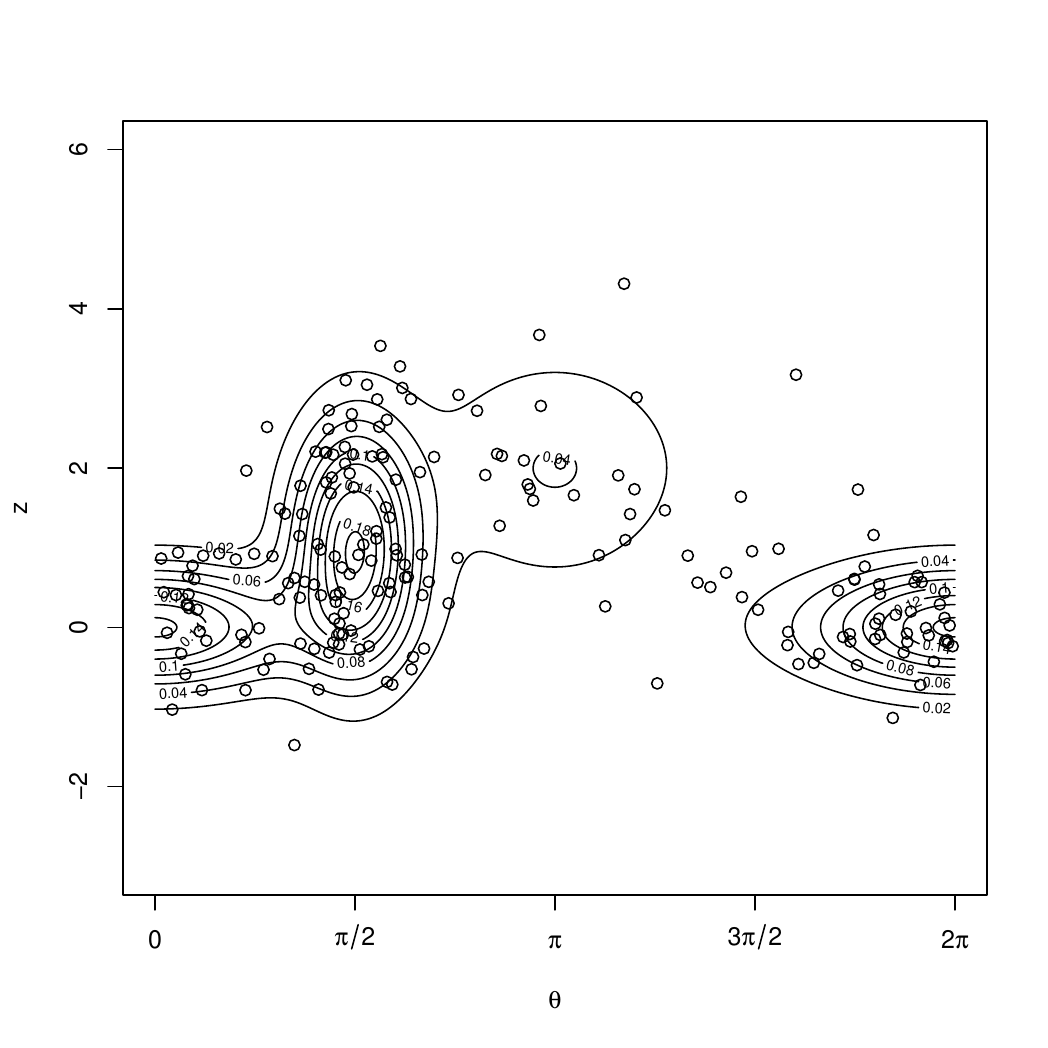}
\includegraphics[width=0.3\textwidth]{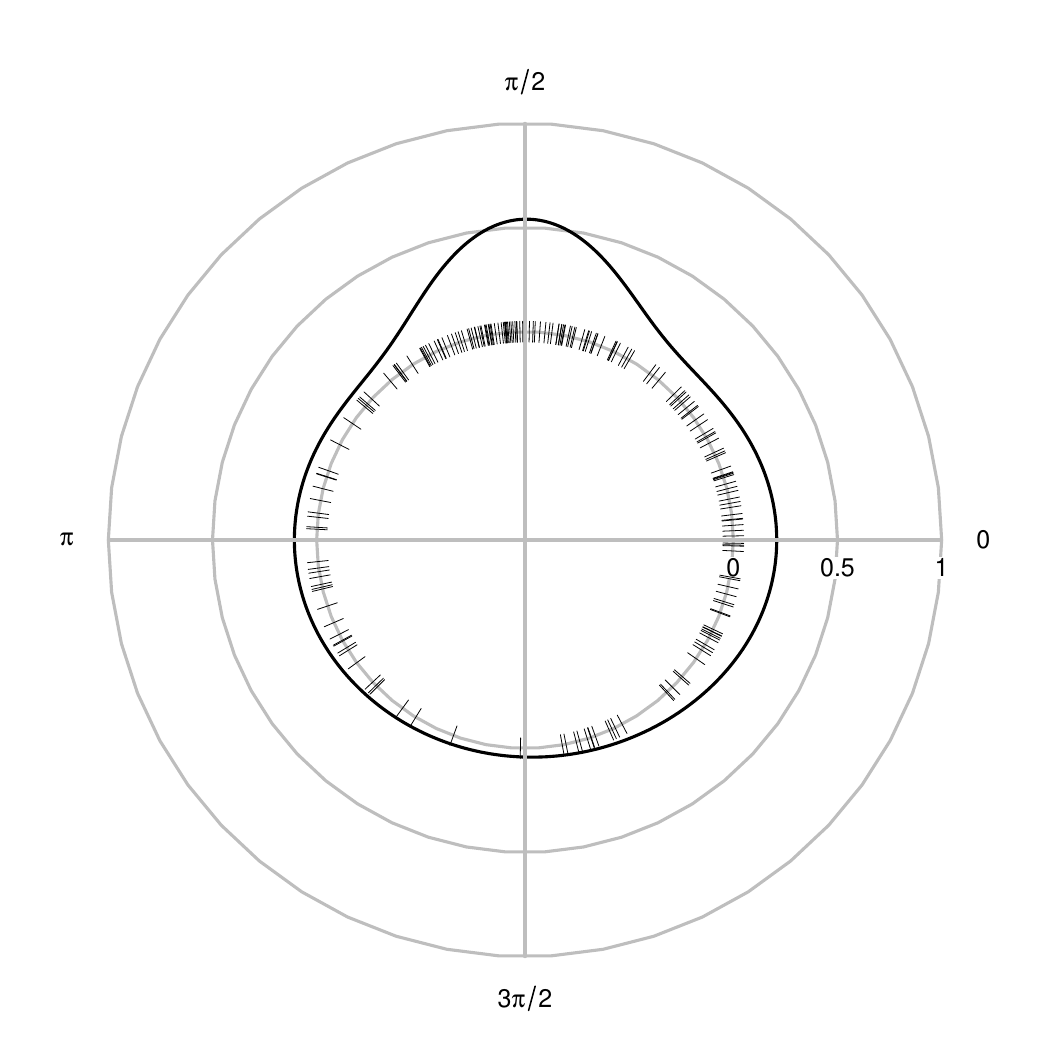}
\includegraphics[width=0.3\textwidth]{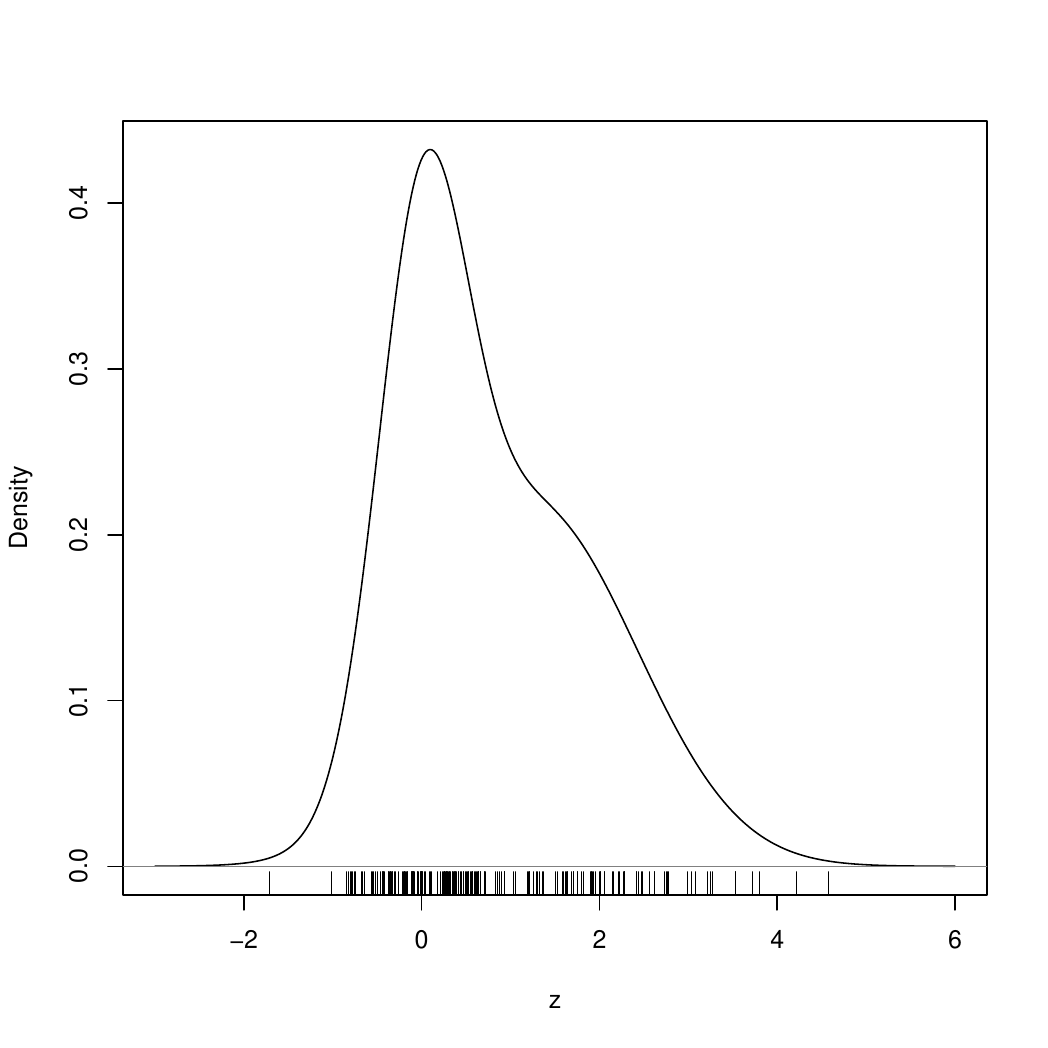}
\caption{\small From left to right: circular-linear mixture (\ref{kerdirlin:mixtdirlin}) and corresponding circular and linear marginal densities, respectively. Random samples of size $n=200$ are drawn. \label{kerdirlin:fig1}}
\end{figure}

Similar results can be obtained for the directional and directional-linear estimators, when considering mixtures of von Mises for the directional case, and mixtures of von Mises and normals for the directional-linear scenario (see Figure \ref{kerdirlin:fig1} for some examples). For the directional setting, an $r$-mixture of von Mises with means $\bmu_j$ and concentration parameters $\kappa_j$, for $j=1,\ldots,r$ is given \nolinebreak[4] by
\begin{align}
f_r(\bx)=\sum_{j=1}^r p_j f_{\mathrm{vM}}(\bx;\bmu_j,\kappa_j),\quad \sum_{j=1}^r p_j=1,\quad p_j\geq 0.\label{kerdirlin:mise:mvm}
\end{align}
Consider a random sample $\bX_1,\ldots,\bX_n$, of a directional variable $\bX$ with density $f_r$ (see Figure \ref{kerdirlin:fig0}, right plot). The following result gives a closed expression for the MISE of the directional kernel estimator.

\begin{prop}
\label{kerdirlin:mise:th:1}
Let $f_r$ be the density of an $r$-mixture of directional von Mises (\ref{kerdirlin:mise:mvm}). The exact MISE of the directional kernel estimator (\ref{kerdirlin:kernel_directional}), obtained from a random sample of size $n$, with von Mises kernel $L(r)=e^{-r}$ is
\begin{align}
\mise{\hat f_h}=\lrp{D_q(h)n}^{-1}+\mathbf{p}^T\lrc{(1-n^{-1})\mathbf{\Psi_2}(h)-2\mathbf{\Psi_1}(h)+\mathbf{\Psi_0}(h)}\mathbf{p}, \label{kerdirlin:mise:mvm:1}
\end{align}
where $\mathbf{p}=\lrp{p_1,\ldots,p_r}^T$ and $D_q(h)=C_q\lrp{1/h^2}^2C_q\lrp{2/h^2}^{-1}$. The matrices $\mathbf{\Psi_a}(h)$, $a=0,1,2$ have entries:
\begin{align*}
\mathbf{\Psi_0}(h)=&\,\lrp{\frac{C_q(\kappa_i)C_q(\kappa_j)}{C_q(||\kappa_i\bmu_i+\kappa_j\bmu_j||)}}_{ij},\\
\mathbf{\Psi_1}(h)=&\,C_q(1/h^2) \lrp{C_q(\kappa_i)C_q(\kappa_j)\Iq{\frac{e^{\kappa_j\bx^T\bmu_j}}{C_q\lrp{\norm {\bx/h^2+\kappa_i\bmu_i}}}}{\bx}}_{ij},\\
\mathbf{\Psi_2}(h)=&\,C_q(1/h^2)^2 \lrp{C_q(\kappa_i)C_q(\kappa_j)\Iq{\lrc{C_q(||\bx/h^2+\kappa_i\bmu_i||)C_q(||\bx/h^2+\kappa_j\bmu_j||)}^{-1}}{\bx}}_{ij},
\end{align*}
where $C_q$ is defined in equation (\ref{kerdirlin:dir:vm}).
\end{prop}

The matrices involved in (\ref{kerdirlin:mise:mvm:1}) are not as simple as the ones for the linear case, due to the convolution properties of the von Mises density. For practical implementation of the exact MISE, it should be noticed that matrices $\mathbf{\Psi_2}(h)$ and $\mathbf{\Psi_1}(h)$  can be evaluated using numerical integration in $q$-spherical coordinates. For clarity purposes, constants $C_q(\kappa_i)$ are included inside matrices $\mathbf{\Psi_2}(h)$, $\mathbf{\Psi_1}(h)$ and $\mathbf{\Psi_0}(h)$ but it is computationally more efficient to consider them within the weights, that is, take $\mathbf{p}=\lrp{p_1C_q(\kappa_1),\ldots,p_rC_q(\kappa_r)}$.

From Proposition \ref{kerdirlin:mise:th:1}, it is easy to derive an analogous result for the case of a $r$-mixture of directional-linear independent von Mises and normals:
\begin{align}
f_r(\bx,z)=\sum_{j=1}^r p_j f_{\mathrm{vM}}(\bx;\bmu_j,\kappa_j)\times\phi_{\sigma_j}\lrp{z-m_j},\quad \sum_{j=1}^r p_j=1,\quad p_j\geq 0.\label{kerdirlin:mise:mvm:3}
\end{align}

\begin{prop}
\label{kerdirlin:mise:th:2}
Let $f_r$ be the density of an $r$-mixture of directional-linear independent von Mises and normals densities given in (\ref{kerdirlin:mise:mvm:3}). For a random sample of size $n$, the exact MISE of the directional-linear kernel density estimator (\ref{kerdirlin:kernel_dirlinear}) with von Mises-normal kernel $LK(r,t)=e^{-r}\times\phi_1(t)$ \nolinebreak[4]is
\begin{align*}
\mise{\hat f_{h,g}}=&\,\lrp{D_q(h)2\pi^\frac{1}{2}gn}^{-1}\\
&+\mathbf{p}^T\lrc{(1-n^{-1})\mathbf{\Psi_2}(h)\circ\mathbf{\Omega_2}(g)-2\mathbf{\Psi_1}(h)\circ\mathbf{\Omega_1}(g)+\mathbf{\Psi_0}(h)\circ\mathbf{\Omega_0}(g)}\mathbf{p},
\end{align*}
where $\circ$ denotes the Hadamard product between matrices and the involved terms are defined as in Proposition \ref{kerdirlin:mise:th:1} and equation (\ref{kerdirlin:mise:no}).
\end{prop}

Once the exact MISE and the AMISE for mixtures of von Mises and normals are derived, it is possible to compare these two error criteria. To that end, let consider the following directional mixture
\begin{align}
\frac{2}{5}\mathrm{vM}\lrp{(1,\mathbf{0}_q)),2}+\frac{2}{5}\mathrm{vM}\lrp{(\mathbf{0}_q,1),10}+\frac{1}{5}\mathrm{vM}\lrp{(-1,\mathbf{0}_q),2},\label{kerdirlin:mixtdir}
\end{align}
where $\mathbf{0}_q$ represents a vector of $q$ zeros, and the directional-linear mixture
\begin{align}
\frac{2}{5}\mathcal{N}\lrp{0,\frac{1}{4}}\times \mathrm{vM}\lrp{(1,\mathbf{0}_q)),2}&+\frac{2}{5}\mathcal{N}\lrp{1,1}\times \mathrm{vM}\lrp{(\mathbf{0}_q,1),10}\nonumber\\
&+\frac{1}{5}\mathcal{N}\lrp{2,1}\times \mathrm{vM}\lrp{(-1,\mathbf{0}_q),2}.\label{kerdirlin:mixtdirlin}
\end{align}

\begin{figure}[h!]
	\centering
	\includegraphics[width=0.3\textwidth]{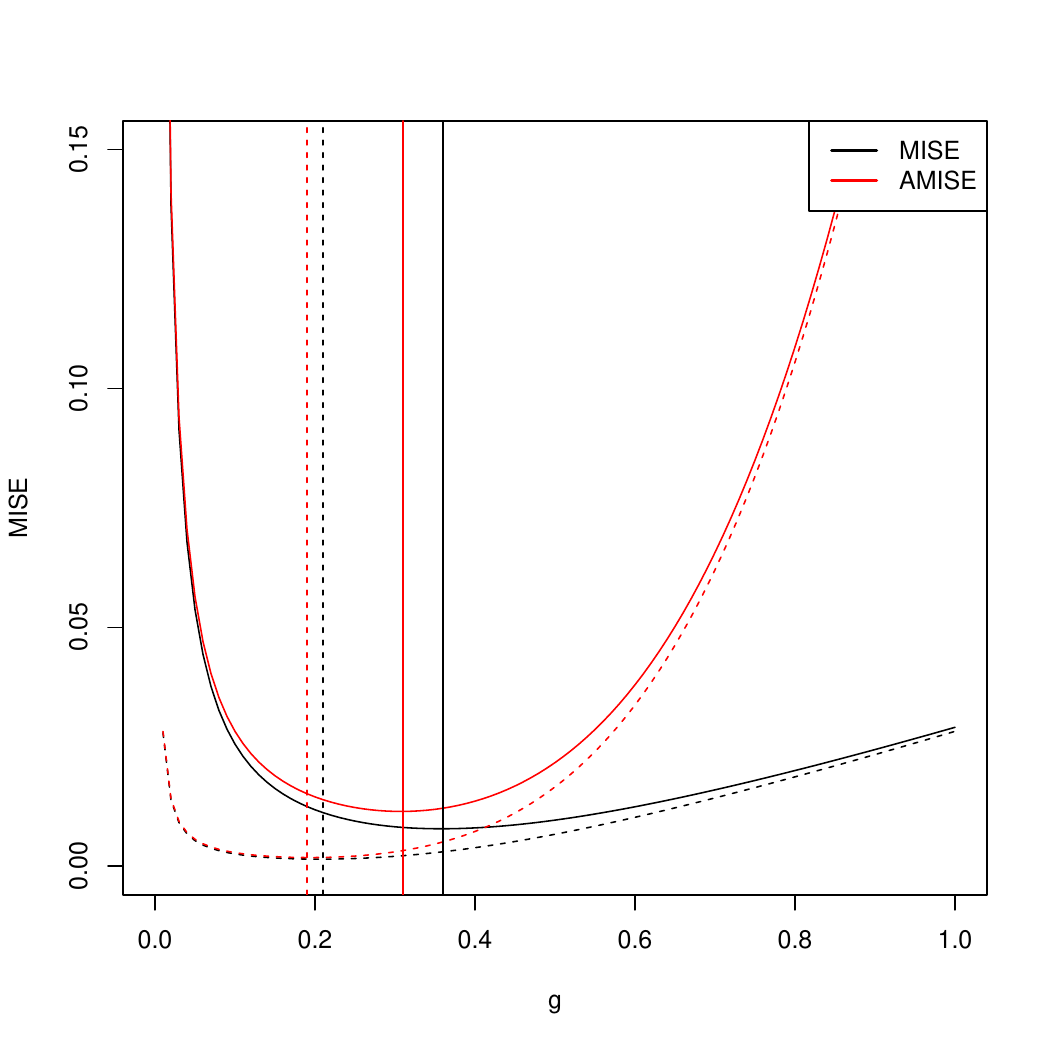}
	\includegraphics[width=0.3\textwidth]{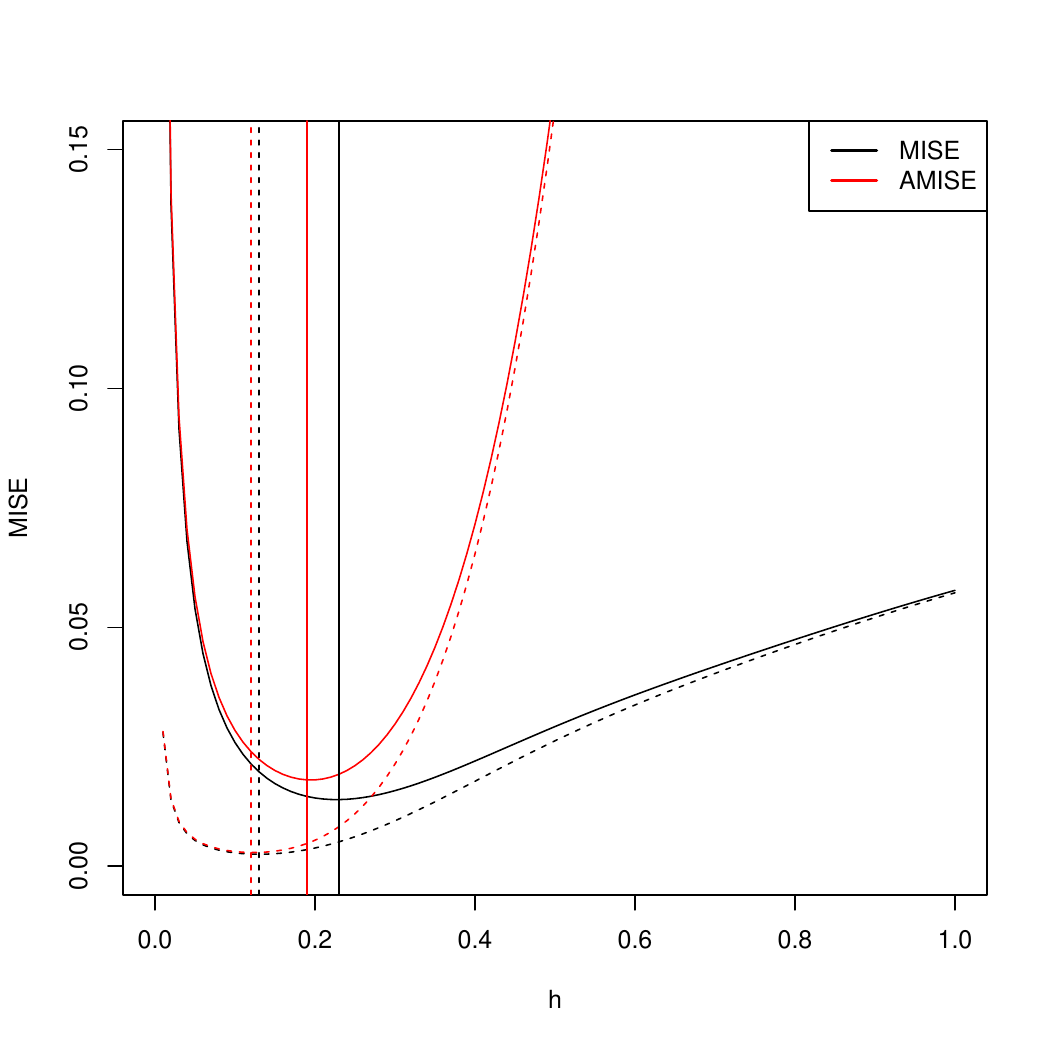}
	\includegraphics[width=0.3\textwidth]{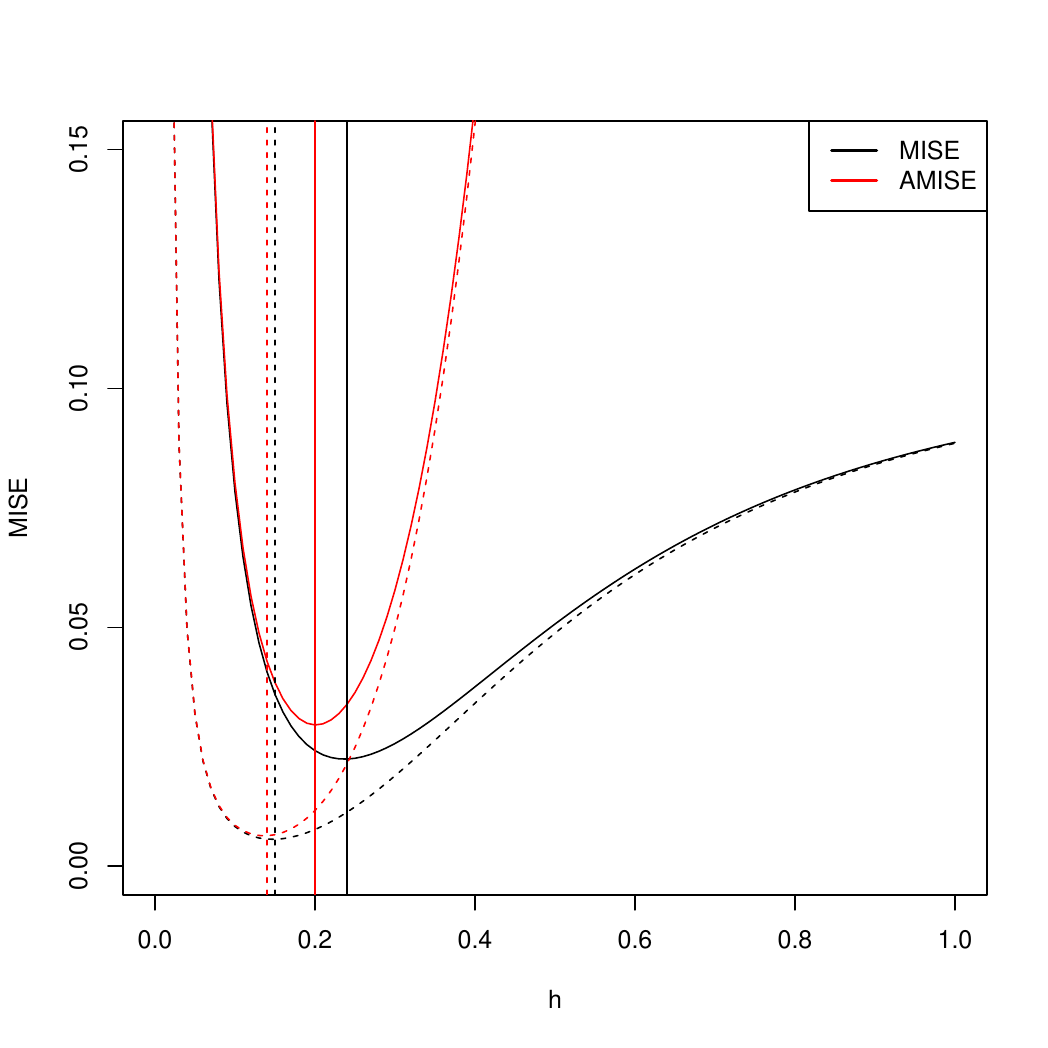}
	\caption{\small From left to right: exact MISE and AMISE for the linear mixture $\frac{2}{5}\mathcal{N}\lrp{0,\frac{1}{4}}+\frac{2}{5}\mathcal{N}\lrp{1,1}+\frac{1}{5}\mathcal{N}\lrp{2,1}$ and the circular and spherical mixtures (\ref{kerdirlin:mixtdir}), for a range of bandwidths between $0$ and $1$. The black curves are for the MISE, whereas the red ones are for the AMISE. Solid curves correspond to $n=100$ and dotted to $n=1000$. Vertical lines represent the bandwidth values minimizing each curve. \label{kerdirlin:fig2}}
\end{figure}

Figure \ref{kerdirlin:fig2} shows the comparison between the exact and asymptotic MISE for the linear, circular and spherical case. As first noted by \cite{Marron1992} for the linear estimator, there exist significative differences between these two errors, being the most remarkable one the rapid growth of the AMISE with respect to the MISE for larger values of the bandwidth. This effect is due to the fact that, for a general bandwidth $h$, $\lim_{h\to\infty}\mathrm{AMISE}\big[\hat f_h\big]=\infty$ since $\mathrm{AMISE}\big[\hat f_h\big]$ is proportional to $h^4$, whereas the MISE level offs at $\lim_{h\to\infty}\mathrm{MISE}\big[\hat f_h\big]=\Iq{f(\bx)^2}{\bx}$. Besides, for the directional case, this effect seems to be augmented probably because of a scale effect in the bandwidths, in the sense that the support of the directional variables is bounded, which is not the case for the linear ones considered. However, although the AMISE and MISE curves differ significantly, the corresponding optimal bandwidths get closer for increasing sample sizes.

\begin{figure}[H]
	\centering
	\includegraphics[scale=0.4]{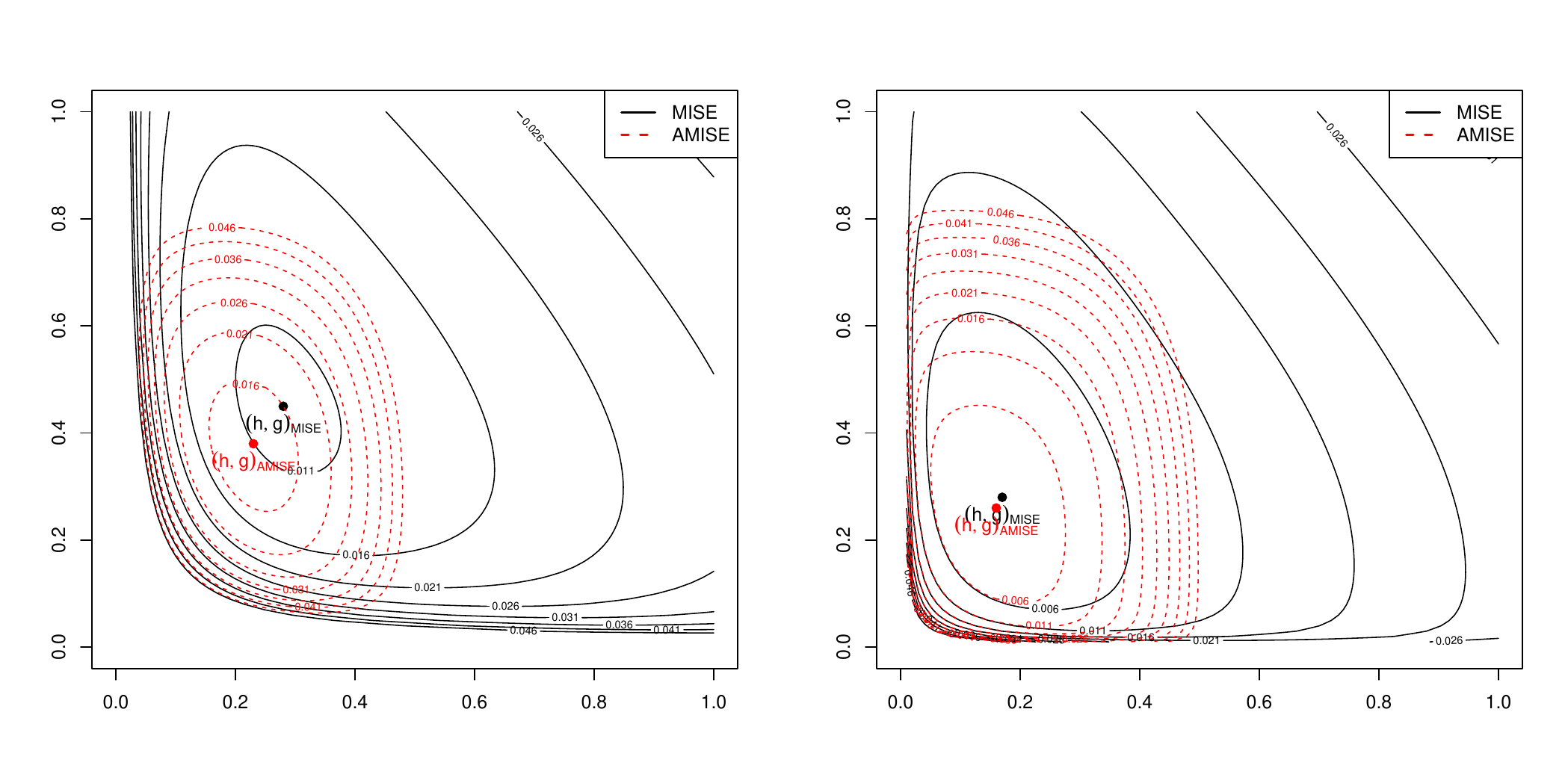}\\
	\includegraphics[scale=0.4]{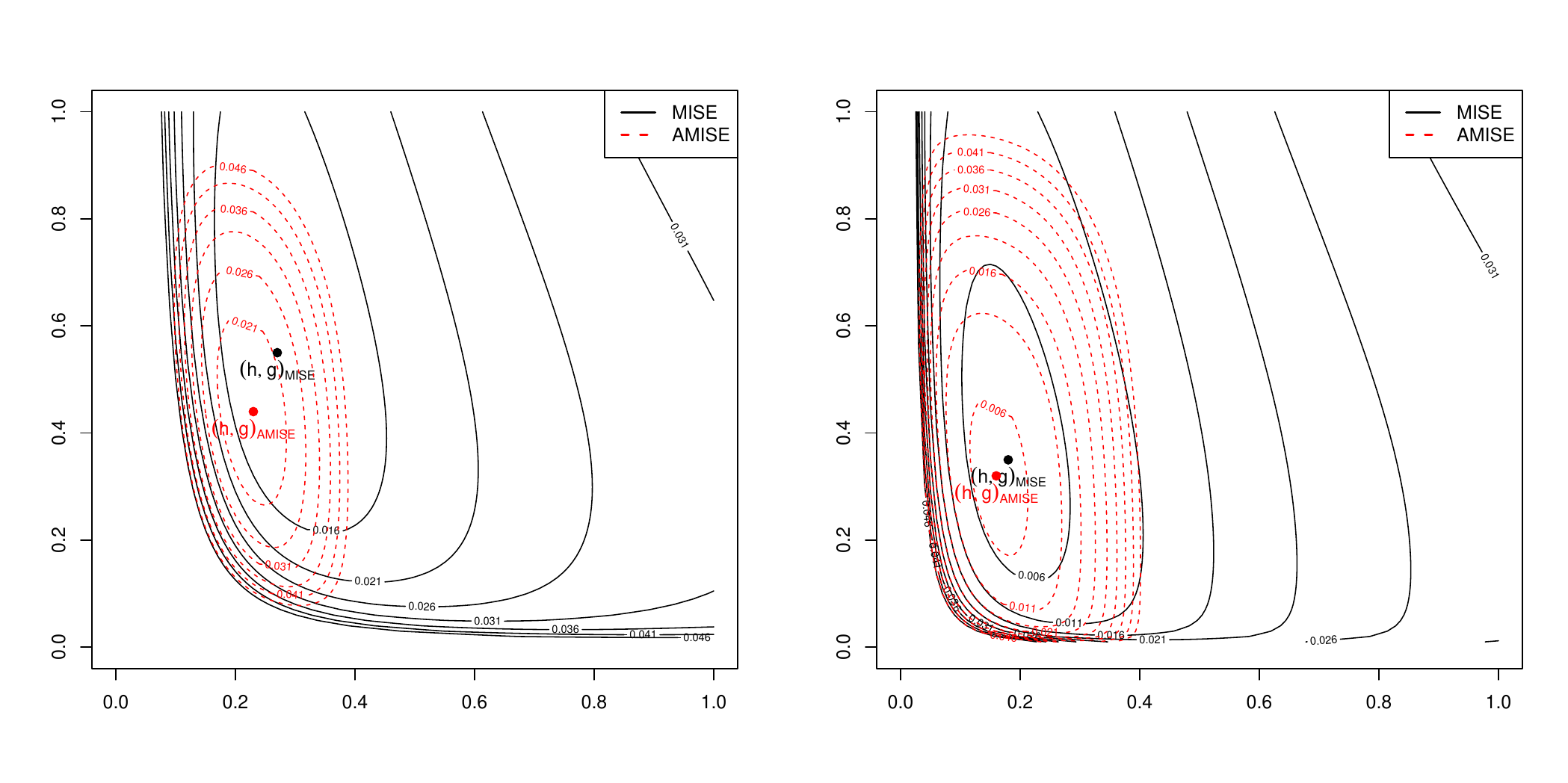}
	\caption{\small Upper plot, from left to right: exact MISE versus AMISE for the circular-linear mixture (\ref{kerdirlin:mixtdirlin}) for $n=100$ and $n=1000$. Lower plot, from left to right: spherical-linear mixture (\ref{kerdirlin:mixtdirlin}) for $n=100$ and $n=1000$. The solid curves are for the MISE, where the dashed ones are for the AMISE. The pairs of bandwidths that minimizes each surface error are denoted by $(h,g)_\mathrm{MISE}$ and by $(h,g)_\mathrm{AMISE}$. \label{kerdirlin:fig3}}
\end{figure}

Figure \ref{kerdirlin:fig3} contains the contourplots of the exact and asymptotic MISE for the circular-linear and spherical-linear cases. The conclusions are more or less the same as for Figure \ref{kerdirlin:fig2}: the asymptotic MISE grows rather quickly than the exact MISE for large values of $h$ or $g$. On the other hand, the contour lines of both surfaces are quite close for small values of the bandwidths and the optimal\nopagebreak[4] bandwidths also get closer for larger sample sizes.

As an immediate application of Propositions \ref{kerdirlin:mise:th:1} and \ref{kerdirlin:mise:th:2}, a bootstrap version of the MISE for the directional and directional-linear estimators can be derived. The bootstrap MISE is an estimator of the true MISE obtained by considering a smooth bootstrap resampling scheme, which will be briefly detailed. In the linear case, the bootstrap MISE is given by
\begin{align*}
\mathrm{MISE}^*_{g_P}\lrc{\hat f_g}=\mathbb{E}^*\lrc{\Ir{ \lrp{\hat f_{g}^*(z)-\hat f_{g_P}(z)}^2}{z}},
\end{align*}
where $\hat f^*_g(z)=\frac{1}{ng}\sum_{i=1}^nK\lrp{\frac{z-Z_i^\ast}{g}}$, being the sample $Z_1^\ast,\ldots,Z_n^\ast$ distributed as $\hat f_{g_P}$. In this case, $g_P$ is a pilot bandwidth and the expectation $\mathbb{E}^*$ is taken with respect to the density estimator $\hat f_{g_P}$. For the linear case, \cite{Cao1993} derived an exact closed expression for $\mathrm{MISE}^*_{g_P}\big[\hat f_g\big]$ that actually avoids the needing of resampling and obtained a bandwidth that minimizes the bootstrap MISE by\nopagebreak[4] previously computing a suitable pilot bandwidth $g_P$. \\

The following two results show the bootstrap MISE expressions for the estimators (\ref{kerdirlin:kernel_directional}) and (\ref{kerdirlin:kernel_dirlinear}) in the case where the kernels are von Mises and normals. As in the linear case, no resampling is needed for computing the bootstrap MISE. These bootstrap versions of the error provide an overall summary of the estimator behaviour, with no restriction on the underlying densities, as long as von Mises and normal kernels are considered. In addition, the following results could be used to derive a bandwidth selector, but it will depend on the selection of pilot bandwidths for both components, which is not an easy problem. 

\begin{coro}
\label{kerdirlin:dir:cor:2}
The bootstrap MISE for directional data, given a sample of length $n$, the von Mises kernel $L(r)=e^{-r}$ and a pilot bandwidth $h_P$, is:
\begin{align*}
\mathrm{MISE}_{h_P}^\ast\lrc{\hat f_h}=\lrp{D_q(h)n}^{-1}+n^{-2}\mathbf{1}^T\lrc{(1-n^{-1})\mathbf{\Psi^\ast_2}(h)-2\mathbf{\Psi^\ast_1}(h)+\mathbf{\Psi^\ast_0}(h)}\mathbf{1},
\end{align*}
where the matrices $\mathbf{\Psi^\ast_a}(h)$, $a=0,1,2$ have the same entries as $\mathbf{\Psi_a}(h)$ but with $\kappa_i=1/h_P^2$ and $\bmu_i=\bX_i$ for $i=1,\ldots,n$. 
\end{coro}

\begin{rem}
The particular case where $q=1$ and $h_P=h$, Corollary \ref{kerdirlin:dir:cor:2} corresponds to the expression of the bootstrap MISE given in \cite{DiMarzio2011}.
\end{rem}

\begin{coro}
\label{kerdirlin:dirlin:cor:2}
The bootstrap MISE for directional-linear data, given a sample of length $n$, the von Mises-normal kernel $LK(r,t)=e^{-r}\times\phi_1(t)$ and a pair of pilot bandwidths $(h_P,g_P)$, is:
\begin{align*}
\mathrm{MISE}_{h_P,g_P}^\ast\lrc{\hat f_{h,g}}=&\,\lrp{D_q(h)2\pi^\frac{1}{2}gn}^{-1}\\
&+n^{-2}\mathbf{1}^T\lrc{(1-n^{-1})\mathbf{\Psi_2^\ast}(h)\circ\mathbf{\Omega_2^\ast}(g)-2\mathbf{\Psi_1^\ast}(h)\circ\mathbf{\Omega_1^\ast}(g)+\mathbf{\mathbf{\Psi_0}^\ast}(h)\circ\mathbf{\Omega_0^\ast}(g)}\mathbf{1},
\end{align*}
where the matrices $\mathbf{\Psi^\ast_a}(h)$ and $\mathbf{\Omega^\ast_a}(g)$, $a=0,1,2$ have the same entries as $\mathbf{\Psi_a}(h)$ and $\mathbf{\Omega_a}(g)$ but with $\kappa_i=1/h_P^2$, $\bmu_i=\bX_i$, $m_i=Z_i$ and $\sigma_i=g_P$ for $i=1,\ldots,n$.
\end{coro}

\section{Conclusions}
\label{kerdirlin:sec:conclusions}

A kernel density estimator for directional-linear data is proposed. Bias, variance and asymptotic normality of the estimator are derived, as well as expressions for the MISE and AMISE. For the particular case of mixtures of von Mises, for directional data, and mixtures of von Mises and normals, in the directional-linear case, the exact expressions for the MISE are obtained, which enables\nopagebreak[4] the comparison with their asymptotic versions. \\

Undoubtedly, one of the main issues in kernel estimation is the appropriate selection of the bandwidth parameter. Although an optimal pair of bandwidths in the AMISE sense has been derived, further research must be done in order to obtain a bandwidth selection method that could be applied in practice. This problem extends somehow to the directional setting, where (likelihood and least squares) cross-validation methods seem to be the available procedures. However, the exact MISE computations open a route to develop bandwidth selectors, for instance, following the ideas in \cite{Oliveira2012}. In fact, a bootstrap version for the MISE when assuming that the underlying mode is a mixture allows for the derivation of bootstrap bandwidths, as in \cite{Cao1993} for the linear\nolinebreak[4] case.\\

A straightforward extension of the proposed estimator can be found in the directional-multidimen\-sional setting, considering a multidimensional random variable. In this case, the linear part of the estimator should be properly adapted including a multidimensional kernel and possibly a bandwidth\nolinebreak[4] matrix.

\section*{Acknowledgements}

The authors acknowledge the support of Project MTM2008-03010, from the Spanish Ministry of Science and Innovation, Project 10MDS207015PR from Direcci\'on Xeral de I+D, Xunta de Galicia and IAP network StUDyS, from Belgian Science Policy. Work of E. Garc\'ia-Portugu\'es has been supported by FPU grant AP2010-0957 from the Spanish Ministry of Education. The authors also acknowledge the suggestions by two anonymous referees that helped improving this paper.


\appendix

\section{Some technical lemmas}
\label{kerdirlin:appendix:techlemmas}

Some technical lemmas that will be used along the proofs of the main results are introduced in this section. To begin with, Lemma \ref{kerdirlin:dir:lem:1a} establishes the asymptotic behaviour of $\lambda_{h,q}(L)$ in (\ref{kerdirlin:normalizing}). With the aim of clarifying the computation of the integrals in the proofs of the main results, Lemma \ref{kerdirlin:dir:lem:1} details a change of variables in $\Om{q}$, whereas Lemma \ref{kerdirlin:dir:lem:2} is used to simplify integrals in $\Om{q}$. Lemma \ref{kerdirlin:dir:lem:3} shows some of the constants introduced along the work for the case where the kernel is von Mises and, finally, Lemma \ref{kerdirlin:dir:lem:4} states the Lemma 2 of \cite{Zhao2001}.\\

Detailed proofs of these lemmas can be found in Appendix \ref{kerdirlin:appendix:proofstechlemmas}. This appendix also includes a rebuild of the proof of the Lemma \ref{kerdirlin:dir:lem:4}, using the same techniques as for the other results, which presents some differences from the original proof.

\begin{lem}
\label{kerdirlin:dir:lem:1a}
Under condition \ref{kerdirlin:cond:d2}, the limit of $\lambda_{h,q}(L)=\om{q-1}\int_0^{2h^{-2}} L(r) r^{\frac{q}{2}-1}(2-rh^2)^{\frac{q}{2}-1}\,dr$, when $h\to0$, is
\begin{align}
\label{kerdirlin:dir:lem:1a:1}
\lim_{h\to0} \lambda_{h,q}(L)=\lambda_q(L)=2^{\frac{q}{2}-1}\om{q-1}\int_0^{\infty} L(r) r^{\frac{q}{2}-1}\,dr,
\end{align}
where $\om{q}$ is the surface area of $\Om{q}$, for $q\geq1$.
\end{lem}

\begin{lem}[A change of variables in $\Om{q}$]
\label{kerdirlin:dir:lem:1}
Let $f$ be a function defined in $\Om{q}$ and $\by\in\Om{q}$ a fixed point. The integral $\Iq{f(\bx)}{\bx}$ can be expressed in one of the following equivalent integrals:
\begin{align}
\Iq{f(\bx)}{\bx}=&\,\int_{-1}^{1}\int_{\Omega_{q-1}} f\lrp{t,(1-t^2)^\frac{1}{2}\bxi} (1-t^2)^{\frac{q}{2}-1}\,\omega_{q-1}(d\bxi)\,dt\label{kerdirlin:dir:lem:1:2}\\
=&\,\int_{-1}^{1}\int_{\Omega_{q-1}} f\lrp{t\by+(1-t^2)^\frac{1}{2}\bB_{\by}\bxi} (1-t^2)^{\frac{q}{2}-1}\,\omega_{q-1}(d\bxi)\,dt\label{kerdirlin:dir:lem:1:3},
\end{align}
where $\bB_{\by}=\lrp{\bb_1,\ldots,\bb_{q}}_{(q+1)\times q}$ is the semi-orthonormal matrix ($\bB_{\by}^T \bB_{\by}=\bI_{q}$ and $\bB_{\by} \bB_{\by}^T=\bI_{q+1}-\by\by^T$) resulting from the completion of $\by$ to the orthonormal basis $\lrb{\by,\bb_1,\ldots,\bb_{q}}$.
\end{lem}

\begin{lem}
\label{kerdirlin:dir:lem:2}
Consider $\bx\in\Omega_q$, a point in the $q$-dimensional sphere with entries $(x_1,\ldots,x_{q+1})$. For all $i,j=1,\ldots,q+1$, it holds that
\begin{align*}
\Iq{x_i}{\bx}=0,\quad\Iq{x_ix_j}{\bx}=\lb\begin{array}{ll}0,&i\neq j,\\\frac{\om{q}}{q+1},&i=j,\end{array}\ri
\end{align*}
where $\om{q}$ is the surface area of $\Om{q}$, for $q\geq1$.
\end{lem}

\begin{lem}
\label{kerdirlin:dir:lem:3}
For the von Mises kernel, \textit{i.e.}, $L(r)=e^{-r}$, $r\geq0$,
\begin{align*}
c_{h,q}(L)=&\,e^{1/h^2}h^{q-1}(2\pi)^\frac{q+1}{2}\mathcal{I}_\frac{q-1}{2}(1/h^2),\quad \lambda_{q}(L)=(2\pi)^\frac{q}{2},\quad b_q(L)=\frac{q}{2},\quad d_q(L)=2^{-\frac{q}{2}}.
\end{align*}
\end{lem}

\begin{lem}[Lemma 2 in \cite{Zhao2001}]
\label{kerdirlin:dir:lem:4}
Under the conditions \ref{kerdirlin:cond:d1}--\ref{kerdirlin:cond:d3}, the expected value of the directional kernel density estimator in a point $\bx\in\Om{q}$, is
\begin{align*}
\E{\hat f_h(\bx)}=f(\bx)+b_q(L)\Psi(f,\bx)h^2+\order{h^2},
\end{align*}
where $\Psi(f,\bx)$ and $b_q(L)$ are given in (\ref{kerdirlin:Psi_dir}) and (\ref{kerdirlin:bq}), respectively.
\end{lem}

\section{Proofs of the main results}
\label{kerdirlin:appendix:mainresults}

\begin{proof}[Proof of Proposition \ref{kerdirlin:dir:prop:2}]
The variance can be decomposed in two terms as follows:
\begin{align}
\V{\hat f_h(\bx)}=&\,%
\frac{c_{h,q}(L)^2}{n}\E{L^2\lrp{\frac{1-\bx^T\bX}{h^2}}}-n^{-1}\E{\hat f_h(\bx)}^2, \label{kerdirlin:dir:prop:2:proof:1}
\end{align}
where the calculus of the first term is quite similar to the calculus of the bias given in Lemma \ref{kerdirlin:dir:lem:4} and the second is given by the same result.\\

Therefore, analogously to the equation (\ref{kerdirlin:dir:prop:1:proof:1}) of Lemma \ref{kerdirlin:dir:lem:4}, the first addend can be expressed\nolinebreak[4] as
\begin{align}
\frac{c_{h,q}(L)^2}{n}h^q\int_{0}^{2h^{-2}}\!\!\!\!\!L^2(r)r^{\frac{q}{2}-1}(2-h^2r)^{\frac{q}{2}-1}\int_{\Omega_{q-1}} \!\!\!\!\!f\lrp{\bx+\ba_{\bx,\bxi}}\,\omega_{q-1}(d\bxi) \,dr,\label{kerdirlin:dir:prop:2:proof:2}
\end{align}
just replacing the kernel $L$ by the squared kernel $L^2$ and where $\ba_{\bx,\bxi}=-rh^2\bx+h\lrc{r(2-h^2r)}^{\frac{1}{2}}\allowbreak\bB_{\bx}\bxi$ $\in\Om{q}$ with $\bB_{\bx}$ defined as in Lemma \ref{kerdirlin:dir:lem:1}. By condition \ref{kerdirlin:cond:d1}, the Taylor expansion of $f$ at $\bx$ is
\begin{align*}
f(\bx+\ba_{\bx,\bxi})-f(\bx)=\ba_{\bx,\bxi}^T\bnab f(\bx)+\frac{1}{2}\ba_{\bx,\bxi}^T\bHcal f(\bx)\ba_{\bx,\bxi}+\order{\ba_{\bx,\bxi}^T\ba_{\bx,\bxi}}.
\end{align*}
Hence,
\begin{align}
(\ref{kerdirlin:dir:prop:2:proof:2})=&\,\frac{c_{h,q}(L)^2}{n}h^q\int_{0}^{2h^{-2}}\!\!\!\!\!L^2(r)r^{\frac{q}{2}-1}(2-h^2r)^{\frac{q}{2}-1}\Bigg\{f(\bx)-rh^2 \om{q-1} \bx^T\bnab f(\bx)\nonumber\\
&\!+\frac{r^2h^4\om{q-1}}{2} \bx^T\bHcal f(\bx)\bx\!+\frac{h^2 r(2-h^2r)\om{q-1}}{2q}\lrp{\nabla^2 f(\bx)-\bx^T \bHcal f(\bx)\bx}+r\om{q-1}\order{h^2}\!\!\Bigg\}\,dr\nonumber\\
=&\,\frac{c_{h,q}(L)}{n}\lb\om{q-1}\lrc{\int_{0}^{2h^{-2}}c_{h,q}(L)h^qL^2(r)r^{\frac{q}{2}-1}(2-h^2r)^{\frac{q}{2}-1}\,dr}f(\bx)\ri\nonumber\\
&-h^2\om{q-1}\lrc{\int_{0}^{2h^{-2}}c_{h,q}(L)h^{q}L^2(r)r^{\frac{q}{2}}(2-h^2r)^{\frac{q}{2}-1}\,dr}\bx^T\bnab f(\bx)\nonumber\\
&+\frac{h^4\om{q-1}}{2}\lrc{\int_{0}^{2h^{-2}}c_{h,q}(L)h^{q}L^2(r)r^{\frac{q}{2}+1}(2-h^2r)^{\frac{q}{2}-1}\,dr}\bx^T \bHcal f(\bx) \bx\nonumber\\
&+\frac{h^2\om{q-1}}{2}\lrc{\int_{0}^{2h^{-2}}c_{h,q}(L)h^{q}L^2(r)r^{\frac{q}{2}}(2-h^2r)^{\frac{q}{2}}\,dr}q^{-1}\lrp{\nabla^2 f(\bx)-\bx^T\bHcal f(\bx) \bx}\nonumber\\
&\lf+\om{q-1}\lrc{\int_{0}^{2h^{-2}}c_{h,q}(L)h^qL^2(r)r^{\frac{q}{2}}(2-h^2r)^{\frac{q}{2}-1}\,dr}\order{h^2}\rb.\label{kerdirlin:dir:prop:2:proof:3}
\end{align}

The integrals in (\ref{kerdirlin:dir:prop:2:proof:3}) can be simplified. For that purpose, define for $h>0$ and indices $i=-1,0,1$, $j=0,1$ the following function:
\begin{align*}
\phi_{h,i,j}(r)=c_{h,q}(L)h^{q}L^2(r) r^{\frac{q}{2}+i}(2-h^2r)^{\frac{q}{2}-j}\mathbbm{1}_{[0,2h^{-2})}(r),\quad r\in[0,\infty).
\end{align*}
As $n\to\infty$, the bandwidth $h\to0$ and the limit of $\phi_{h,i,j}$ is given by
\begin{align*}
\phi_{i,j}(r)=\lim_{h\to0}\phi_{h,i,j}(r)=\lambda_q(L)^{-1}L^2(r) r^{\frac{q}{2}+i} 2^{\frac{q}{2}-j}\mathbbm{1}_{[0,\infty)}(r).
\end{align*}
Applying the Dominated Convergence Theorem (DCT) and the same techniques of the proof of Lemma \ref{kerdirlin:dir:lem:1a} (see Remark \ref{kerdirlin:rem:phitcd}), it can be seen that:
\begin{align*}
\lim_{h\to0}\int_0^\infty \phi_{h,i,j}(r) \,dr=\lambda_q(L)^{-1}2^{\frac{q}{2}-j}\int_0^\infty L^2(r) r^{\frac{q}{2}+i}\,dr\stackrel{(\ref{kerdirlin:dir:lem:1a:1})}{=}
\lb\begin{array}{ll}
\frac{2^{1-j}}{\om{q-1}}d_q(L),&i=-1,\\
\frac{2^{1-j}}{\om{q-1}}e_q(L),&i=0,\\
\frac{2^{1-j}}{\om{q-1}}\frac{\int_0^\infty L^2(r)r^{\frac{q}{2}+1}\,dr}{\int_0^\infty L(r)r^{\frac{q}{2}-1}\,dr},&i=1,
\end{array}\ri
\end{align*}
where $e_q(L)=\int_0^\infty L^2(r) r^{\frac{q}{2}}\,dr\big/\int_0^\infty L(r) r^{\frac{q}{2}-1}\,dr$. Then, taking into account that $\int_0^\infty \varphi_{h,i,j}(r) \,dr=\int_0^\infty \varphi_{i,j}(r) \,dr\lrp{1+\order{1}}$ the integrals in brackets of (\ref{kerdirlin:dir:prop:2:proof:3}) can be replaced, obtaining that
\begin{align}
(\ref{kerdirlin:dir:prop:2:proof:3})=&\,\frac{c_{h,q}(L)}{n}\lrc{d_q(L)f(\bx)+e_q(L)h^2\Psi(f,\bx)}+\order{(nh^q)^{-1}}.
\label{kerdirlin:dir:prop:2:proof:4}
\end{align}
The second term in (\ref{kerdirlin:dir:prop:2:proof:1}) is given by
\begin{align}
\E{\hat f_h(\bx)}^2 %
=&\,\lrc{f(\bx)+b_q(L)h^2\Psi(f,\bx)}^2+\order{h^2}.
\label{kerdirlin:dir:prop:2:proof:5}
\end{align}

The result holds from (\ref{kerdirlin:dir:prop:2:proof:4}) and (\ref{kerdirlin:dir:prop:2:proof:5}):
\begin{align*}
\V{\hat f_h(\bx)}=&\,\frac{c_{h,q}(L)}{n}\lrc{d_q(L)f(\bx)+e_q(L)h^2\Psi(f,\bx)}\\
&-\frac{1}{n}\lrc{f(\bx)+b_q(L)h^2\Psi(f,\bx)}^2+\order{(nh^q)^{-1}},
\end{align*}
which can be simplified into
\begin{align*}
\V{\hat f_h(\bx)}=&\,\frac{c_{h,q}(L)}{n} d_q(L)f(\bx)+\order{(nh^q)^{-1}}.
\end{align*}
\end{proof}


\begin{proof}[Proof of Proposition \ref{kerdirlin:dirlin:prop:1}]
Denote by $\mathrm{Bias}\big[\hat f_{h,g}(\bx,z)\big]=\mathbb{E}\big[\hat f_{h,g}(\bx,z)\big]-f(\bx,z)$ the bias of the kernel estimator. Applying the change of variables stated in Lemma \ref{kerdirlin:dir:lem:1} and then an ordinary change of variables given by $r=\frac{1-t}{h^2}$, the bias results in:
\begin{align}
\mathrm{Bias}\lrc{\hat f_{h,g}(\bx,z)}=&\,\frac{c_{h,q}(L)}{g}\E{LK\lp\frac{1-\bx^T \bX}{h^2},\frac{z-Z}{g}\rp}-f(\bx,z)\nonumber\\
=&\,\frac{c_{h,q}(L)}{g} \int_{\Omega_q}\int_\R LK\lp\frac{1-\bx^T \by}{h^2},\frac{z-t}{g}\rp \lrp{f(\by,t)-f(\bx,z)}\,dt\,\omega_q(d\by)\nonumber\\
=&\,c_{h,q}(L) \int_{\Omega_q}\int_\R LK\lp\frac{1-\bx^T \by}{h^2}, v\rp \lrp{f(\by,z-gv)-f(\bx,z)}\,dv\,\omega_q(d\by)\nonumber\\
=&\,c_{h,q}(L) \!\int_{-1}^1\int_{\Omega_{q-1}}\!\int_\R LK\lp\frac{1-u}{h^2}, v\rp \Big( f\lrp{u\bx+(1-u^2)^{\frac{1}{2}}\bB_{\bx}\bxi,z-gv}\!-\!f(\bx,z)\Big)\nonumber\\
&\times(1-u^2)^{\frac{q}{2}-1}\,dv\,\omega_{q-1}(d\bxi)\,du\nonumber\\
=&\,c_{h,q}(L) h^q\int_{0}^{2h^{-2}}\int_{\Omega_{q-1}}\int_\R LK\lp r, v\rp \lrp{f\lp (\bx,z)+\ba_{\bx,z,\bxi}\rp-f(\bx,z)} \, dv\,\omega_{q-1}(d\bxi)\nonumber\\
&\times r^{\frac{q}{2}-1}(2-h^2r)^{\frac{q}{2}-1}\,dr\nonumber\\
=&\,c_{h,q}(L) h^q\int_{0}^{2h^{-2}}L\lp r\rp r^{\frac{q}{2}-1}(2-h^2r)^{\frac{q}{2}-1} \int_\R  K\lp v\rp\nonumber\\
&\times\int_{\Omega_{q-1}}\lrp{f\lp (\bx,z)+\ba_{\bx,z,\bxi}\rp-f(\bx,z)}\omega_{q-1}(d\bxi)\,dv\,dr,\label{kerdirlin:dirlin:prop:1:proof:1}
\end{align}
where $\ba_{\bx,z,\bxi}=\Big(-rh^2\bx+h\lc r(2-h^2r)\rc^{\frac{1}{2}}\bB_{\bx}\bxi,-gv\Big) \in \Omega_{q}\times\R$. The computation of the last integral in (\ref{kerdirlin:dirlin:prop:1:proof:1}) is achieved using the multivariate Taylor expansion of $f$ at $(\bx,z)$, in virtue of condition \ref{kerdirlin:cond:dl1}:
\begin{align*}
f((\bx,z)+\ba_{\bx,z,\bxi})-f(\bx,z)=\ba_{\bx,z,\bxi}^T\bnab f(\bx,z)+\frac{1}{2}\ba_{\bx,z,\bxi}^T \bHcal f(\bx,z)\ba_{\bx,z,\bxi}+\order{\ba_{\bx,z,\bxi}^T\ba_{\bx,z,\bxi}}.
\end{align*}
Let denote by $\bga_{\bx,\bxi}=-rh^2\bx+h\lc r(2-h^2r)\rc^{\frac{1}{2}}\bB_{\bx}\bxi$. Bearing in mind the directional and linear components of the gradient $\bnab f(\bx,z)$ and the Hessian matrix $\bHcal f(\bx,z)$, it follows
\begin{align*}
f((\bx,z)+\ba_{\bx,z,\bxi})-f(\bx,z)=&\,\lrc{\bga_{\bx,\bxi}^T\bnab_\bx f(\bx,z)-gv\nabla_z f(\bx,z)}\nonumber\\
&+\frac{1}{2}\lrc{\bga_{\bx,\bxi}^T \bHcal_\bx f(\bx,z)\bga_{\bx,\bxi}-2gv\bga_{\bx,\bxi}^T\bHcal_{\bx,z}f(\bx,z)+g^2v^2\Hcal_z f(\bx,z)}\nonumber\\
&+\order{\ba_{\bx,z,\bxi}^T\ba_{\bx,z,\bxi}}.
\end{align*}
Then, the calculus of the integral $\int_{\Omega_{q-1}}\lrp{f\lp (\bx,z)+\ba_{\bx,z,\bxi}\rp-f(\bx,z)}\,\omega_{q-1}(d\bxi)$ can be split into six addends. 
Second and sixth terms are computed straightforward:
\begin{align}
\int_{\Omega_{q-1}}-gv \nabla_z f(\bx,z)\,\omega_{q-1}(d\bxi)=-\omega_{q-1}\,gv\nabla_z f(\bx,z),\label{kerdirlin:dirlin:prop:1:proof:3}\\
\int_{\Omega_{q-1}} g^2v^2 \Hcal_z f(\bx,z)\,\omega_{q-1}(d\bxi)=\omega_{q-1}\,g^2v^2 \Hcal_z f(\bx,z).\label{kerdirlin:dirlin:prop:1:proof:6}
\end{align}

For the first and fourth addends, by Lemma \ref{kerdirlin:dir:lem:2}, the integration of $\xi_i$ with respect to $\bxi$ is zero:
\begin{align}
\int_{\Omega_{q-1}}\bga_{\bxi,z}^T\bnab_\bx f(\bx,z)\,\omega_{q-1}(d\bxi)=&\,-\omega_{q-1}h^2r\bx^T\bnab_\bx f(\bx,z),\label{kerdirlin:dirlin:prop:1:proof:2}\\
\int_{\Omega_{q-1}}-2gv\bga_{\bx,\bxi} \bHcal_{\bx,z} f(\bx,z)\,\omega_{q-1}(d\bxi)=&\,2gv\omega_{q-1}h^2r\bx^T\bHcal_{\bx,z} f(\bx,z).\label{kerdirlin:dirlin:prop:1:proof:5}
\end{align}

Finally, in the fifth term, the integrand can be decomposed as follows:
\begin{align*}
\bga_{\bx,\bxi}^T \bHcal_\bx f(\bx,z)\bga_{\bx,\bxi}%
=&\,h^4r^2 \bx^T \bHcal_\bx f(\bx,z) \bx+h^2r(2-h^2r) \sum_{i,j =1}^q\xi_i\xi_j\bb_i^T\bHcal_\bx f(\bx,z)\bb_j\\
&-2h^3r^\frac{3}{2}(2-h^2r)^\frac{1}{2}\sum_{i=1}^q\xi_i \bx^T\bHcal_\bx f(\bx,z) \bb_i.
\end{align*}

In virtue of Lemma \ref{kerdirlin:dir:lem:2}, the third addend vanishes as well as the second, except for the diagonal terms. Next, as $\lrb{\bx,\bb_1,\ldots,\bb_{q}}$ is an orthonormal basis in $\R^{q+1}$, the sum of the diagonal terms can be computed by simple algebra:
\begin{align*}
\sum_{i=1}^q \bb_i^T \bHcal_\bx f(\bx,z) \bb_i&=\tr{\bHcal_\bx f(\bx,z) \sum_{i=1}^q \bb_i\bb_i^T}=\tr{\bHcal_\bx f(\bx,z)\lrp{\bI_{q+1}-\bx\bx^T}}\\
&=\nabla_\bx^2 f(\bx,z)-\bx^T\bHcal_\bx f(\bx,z) \bx,
\end{align*}
where $\nabla_\bx^2 f(\bx,z)$ is the Laplacian of $f$ restricted to the directional component $\bx$, $\bI_{q+1}$ is the identity matrix of order $q+1$ and $\mathrm{tr}$ is the trace operator. By Lemma \ref{kerdirlin:dir:lem:2} and the previous calculus, the fifth term is
\begin{align}
\int_{\Omega_{q-1}} &\bga_{\bxi,z}^T \bHcal_\bx f(\bx,z)\bga_{\bxi,z}\,\omega_{q-1}(d\bxi) \nonumber\\
&\;=\omega_{q-1}h^4r^2 \bx^T \bHcal_\bx f(\bx,z) \bx+ \omega_{q-1}h^2r(2-h^2r)q^{-1}\lc\nabla^2_\bx f(\bx,z)-\bx^T\bHcal_\bx f(\bx,z)\bx\rc.\label{kerdirlin:dirlin:prop:1:proof:4}
\end{align}

Note also that the order of $\ba_{\bx,z,\bxi}^T\ba_{\bx,z,\bxi}$ is easily computed:
\begin{align}
\order{\ba_{\bx,z,\bxi}^T\ba_{\bx,z,\bxi}}=&\,r\order{h^2}+v^2\order{g^2}.\label{kerdirlin:dirlin:prop:1:proof:7}
\end{align}

Combining (\ref{kerdirlin:dirlin:prop:1:proof:3})--(\ref{kerdirlin:dirlin:prop:1:proof:7}), and using condition \ref{kerdirlin:cond:dl2} on the kernel $K$:
\begin{align}
(\ref{kerdirlin:dirlin:prop:1:proof:1})=&\,c_{h,q}(L) h^q\int_{0}^{2h^{-2}}L\lp r\rp r^{\frac{q}{2}-1}(2-h^2r)^{\frac{q}{2}-1}\int_\R K\lp v\rp\bigg\{\int_{\Omega_{q-1}} \lrc{\bga_{\bxi,\bx}\bnab_\bx f(\bx,z)-gv\nabla_z f(\bx,z)}\nonumber\\
&+\frac{1}{2}\lrc{\bga_{\bxi,\bx}^T \bHcal_\bx f(\bx,z)\bga_{\bxi,\bx}-2gv\bga_{\bxi,\bx}^T\bHcal_{\bx,z}f(\bx,z)+g^2v^2\Hcal_z f(\bx,z)}\nonumber\\
&+r\order{h^2}+v^2\order{g^2}\,\omega_{q-1}(d\bxi)\bigg\}\,dv\,dr\nonumber\\
=&\,\omega_{q-1}c_{h,q}(L) h^q\int_{0}^{2h^{-2}}L\lp r\rp r^{\frac{q}{2}-1}(2-h^2r)^{\frac{q}{2}-1}\int_\R K(v) \bigg\{-h^2r\bx^T\bnab_\bx f(\bx,z)\nonumber\\
&-gv\nabla_z f(\bx,z)+\frac{1}{2}\lrc{h^4r^2\bx^T \bHcal_\bx f(\bx,z)+h^2 r(2-h^2r)q^{-1}\lp\nabla^2_\bx f(\bx,z)-\bx^T \bHcal_\bx f(\bx,z)\bx\rp}\nonumber\\
&+gvh^2r\bx^T \bHcal_{\bx,z} f(\bx,z)+\frac{g^2v^2}{2} \Hcal_z f(\bx,z)+r\order{h^2}+v^2\order{g^2}\bigg\}\,dv\,dr\nonumber\\
=&\,\omega_{q-1}c_{h,q}(L) h^q\int_{0}^{2h^{-2}}L\lp r\rp r^{\frac{q}{2}-1}(2-h^2r)^{\frac{q}{2}-1}\bigg\{-h^2 r \bx^T\bnab_\bx f(\bx,z)\nonumber\\
&+\frac{1}{2}\Big[h^4r^2\bx^T\bHcal _\bx f(\bx,z)+h^2r(2-h^2r)q^{-1}\lrp{\nabla_\bx^2f(\bx,z)-\bx^T\bHcal _\bx f(\bx,z)\bx}\nonumber\\
&+g^2\bHcal _zf(\bx,z)\mu_2(K)\Big]+r\order{h^2}+\mu_2(K)\order{g^2}\bigg\}\,dr.\label{kerdirlin:dirlin:prop:1:proof:8}
\end{align}

For $h>0$, $i=-1,0,1$, $j=0,1$, consider the following functions
\begin{align*}
\varphi_{h,i,j}(r)=c_{h,q}(L)h^{q}L(r) r^{\frac{q}{2}+i}(2-h^2r)^{\frac{q}{2}-j}\mathbbm{1}_{[0,2h^{-2})}(r),\quad r\in[0,\infty).
\end{align*}
When $n\to\infty$, $h\to0$ and the limit of $\varphi_{h,i,j}$ is given by
\begin{align*}
\varphi_{i,j}(r)=\lim_{h\to0}\varphi_{h,i,j}(r)=\lambda_q(L)^{-1}L(r) r^{\frac{q}{2}+i} 2^{\frac{q}{2}-j}\mathbbm{1}_{[0,\infty)}(r).
\end{align*}
Applying Remark \ref{kerdirlin:rem:phitcd} of Lemma \ref{kerdirlin:dir:lem:1a},
\begin{align*}
\lim_{h\to0}\int_0^\infty \varphi_{i,j,h}(r) \,dr=\lambda_q(L)^{-1}2^{\frac{q}{2}-j}\int_0^\infty L(r) r^{\frac{q}{2}+i}\,dr\stackrel{(\ref{kerdirlin:dir:lem:1a:1})}{=}\lb\begin{array}{ll}
\frac{2^{1-j}}{\om{q-1}},&i=-1,\\
\frac{2^{1-j}}{\om{q-1}}b_q(L),&i=0,\\
\frac{2^{1-j}}{\om{q-1}}\frac{\int_0^\infty L(r)r^{\frac{q}{2}+1}\,dr}{\int_0^\infty L(r)r^{\frac{q}{2}-1}\,dr},&i=1.\\
\end{array}\ri
\end{align*}

Then, the six integrals in (\ref{kerdirlin:dirlin:prop:1:proof:8}) can be written using $\int_0^\infty \varphi_{i,j,h}(r) \,dr=\int_0^\infty \varphi_{i,j}(r) \,dr\lrp{1+\order{1}}$. Replacing this in (\ref{kerdirlin:dirlin:prop:1:proof:8}) leads to
\begin{align*}
(\ref{kerdirlin:dirlin:prop:1:proof:8})=&\,-h^2\om{q-1}\lrc{\frac{b_q(L)}{\om{q-1}}+\order{1}}\bx^T\bnab_\bx f(\bx,z)\nonumber\\
&+\frac{h^4\om{q-1}}{2}\lrc{\frac{b_q(L)}{\om{q-1}}\frac{\int_0^\infty L(r)r^{\frac{q}{2}+1}\,dr}{\int_0^\infty L(r)r^\frac{q}{2}\,dr}+\order{1}} \bx^T\bHcal_\bx f(\bx,z)\bx\nonumber\\
&+\frac{h^2\om{q-1}}{2}\lrc{\frac{b_q(L)}{\om{q-1}}+\order{1}} q^{-1}\lrp{\nabla_\bx^2 f(\bx,z)-\bx^T\bHcal_\bx f(\bx,z) \bx}\nonumber\\
&+\frac{\om{q-1}}{2}\lrc{\frac{1}{\om{q-1}}+\order{1}}g^2\Hcal_z f(\bx,z)\mu_2(K)\nonumber\\
&+\om{q-1}\lrc{\frac{b_q(L)}{\om{q-1}}+\order{1}}\order{h^2}+\om{q-1}\lrc{\frac{1}{\om{q-1}}+\order{1}}\order{g^2}\nonumber\\
=&\,h^2b_q(L)\lrc{-\bx^T\bnab_\bx f(\bx,z)+q^{-1}\lrp{\nabla_\bx^2 f(\bx)-\bx^T\bHcal_\bx f(\bx,z) \bx}}+g^2\Hcal_z f(\bx,z)\mu_2(K)\nonumber\\
&+\Order{h^4}+\order{h^2}+\order{g^2}\nonumber\\
=&\,h^2b_q(L)\Psi_\bx(f,\bx,z)+\frac{g^2}{2}\Hcal_z f(\bx,z)\mu_2(K)+\order{h^2+g^2}.
\end{align*}
\end{proof}


\begin{proof}[Proof of Proposition \ref{kerdirlin:dirlin:prop:2}]

The variance can be decomposed as
\begin{align}
\V{\hat f_{h,g}(\bx,z)}%
=&\,\frac{c_{h,q}(L)^2}{ng^2}\E{LK^2\lrp{\frac{1-\bx^T\bX}{h^2},\frac{z-Z}{g}}}-n^{-1}\E{\hat f_{h,g}(\bx,z)}^2,\label{kerdirlin:dirlin:prop:2:proof:1}
\end{align}
where the calculus of the first term is quite similar to the calculus of the bias and the second is given in the previous result.\\

Analogous to (\ref{kerdirlin:dirlin:prop:1:proof:1}),
\begin{align}
\frac{c_{h,q}(L)^2}{ng^2}\E{LK^2\lrp{\frac{1-\bx^T\bX}{h^2},\frac{z-Z}{g}}}=&\,\frac{c_{h,q}(L)^2}{ng}h^q\int_0^{2h^{-2}}L^2(r)r^{\frac{q}{2}-1}(2-h^2r)^{\frac{q}{2}-1}\int_\R K^2(v)\nonumber\\
&\times\int_{\Om{q-1}}f((\bx,z)+\ba_{\bx,z,\bxi})\,\om{q-1}(d\bxi)\,dv\,dr,\label{kerdirlin:dirlin:prop:2:proof:2}
\end{align}
just replacing $LK$ by $LK^2$. Then, using that $K^2$ is a symmetric function around zero:
\begin{align}
\int_\R K^2(v) \,dv=R(K),\,\int_\R v K^2(v) \,dv=0,\,\int_\R v^2 K^2(v) \,dv=\mu_2\lrp{K^2}\label{kerdirlin:dirlin:prop:2:proof:3},
\end{align}
Applying the multivariate Taylor expansion of $f$ at $(\bx,z)$ and by (\ref{kerdirlin:dirlin:prop:2:proof:3}), equation (\ref{kerdirlin:dirlin:prop:2:proof:2}) results\nolinebreak[4] in
\begin{align}
(\ref{kerdirlin:dirlin:prop:2:proof:2})=&\,\omega_{q-1}\frac{c_{h,q}(L)^2}{ng} h^q\int_{0}^{2h^{-2}}L^2(r) r^{\frac{q}{2}-1}(2-h^2r)^{\frac{q}{2}-1}\int_\R K^2(v) \bigg\{f(\bx,z)-h^2r\bx^T\bnab_\bx f(\bx,z)\nonumber\\
&-gv\nabla_z f(\bx,z)+\frac{1}{2}\Big[h^4r^2\bx^T \bHcal_\bx f(\bx,z)+h^2 r(2-h^2r)q^{-1}\lp\nabla^2_\bx f(\bx,z)-\bx^T \bHcal_\bx f(\bx,z)\bx\rp\Big]\nonumber\\
&+gvh^2r\bx^T \bHcal_{\bx,z} f(\bx,z)+\frac{g^2v^2}{2} \Hcal_z f(\bx,z)+r\order{h^2}+v^2\order{g^2}\bigg\}\,dv\,dr\nonumber\\
\stackrel{\mathclap{(\ref{kerdirlin:dirlin:prop:2:proof:3})}}{=}\,&\,\,\omega_{q-1}\frac{c_{h,q}(L)^2}{ng} h^q\int_{0}^{2h^{-2}}L^2(r) r^{\frac{q}{2}-1}(2-h^2r)^{\frac{q}{2}-1}\bigg\{R(K)f(\bx,z)-R(K)h^2r\bx^T\bnab_\bx f(\bx,z)\nonumber\\
&+\frac{R(K)}{2}\Big[h^4r^2\bx^T \bHcal_\bx f(\bx,z)+h^2 r(2-h^2r)q^{-1}\lp\nabla^2_\bx f(\bx,z)-\bx^T \bHcal_\bx f(\bx,z)\bx\rp\Big]\nonumber\\
&+\mu_2\lrp{K^2}\frac{g^2}{2} \Hcal_z f(\bx,z)+r\order{h^2}+\mu_2(K^2)\order{g^2}\bigg\}\,dr.\label{kerdirlin:dirlin:prop:2:proof:4}
\end{align}

Define the following functions, for $h>0$, $i=-1,0,1$ and $j=0,1$:
\begin{align*}
\phi_{h,i,j}(r)=c_{h,q}(L)h^{q}L^2(r) r^{\frac{q}{2}+i}(2-h^2r)^{\frac{q}{2}-j}\mathbbm{1}_{[0,2h^{-2})}(r),\quad r\in[0,\infty).
\end{align*}
When $n\to\infty$, $h\to0$ and the limit of $\phi_{h,i,j}$ is given by
\begin{align*}
\phi_{i,j}(r)=\lim_{h\to0}\phi_{h,i,j}(r)=\lambda_q(L)^{-1}L^2(r) r^{\frac{q}{2}+i} 2^{\frac{q}{2}-j}\mathbbm{1}_{[0,\infty)}(r).
\end{align*}
Applying the same techniques of the proof of Lemma \ref{kerdirlin:dir:lem:1a} to the functions $\phi_{h,i,j}$ with the different values of $i,j$ and $L^2$ instead of $L$, and using the relation (\ref{kerdirlin:normalizing}), it follows:
\begin{align*}
\lim_{h\to0}\int_0^\infty \phi_{h,i,j}(r) \,dr=\lambda_q(L)^{-1}2^{\frac{q}{2}-j}\int_0^\infty L^2(r) r^{\frac{q}{2}+i}\,dr\stackrel{(\ref{kerdirlin:dir:lem:1a:1})}{=}
\lb\begin{array}{ll}
\frac{2^{1-j}}{\om{q-1}}d_q(L),&i=-1,\\
\frac{2^{1-j}}{\om{q-1}}e_q(L),&i=0,\\
\frac{2^{1-j}}{\om{q-1}}\frac{\int_0^\infty L^2(r)r^{\frac{q}{2}+1}\,dr}{\int_0^\infty L(r)r^{\frac{q}{2}-1}\,dr},&i=1,
\end{array}\ri
\end{align*}
where $e_q(L)=\int_0^\infty L^2(r) r^{\frac{q}{2}}\,dr\big/\int_0^\infty L(r) r^{\frac{q}{2}-1}\,dr$. So, for the terms between square brackets of (\ref{kerdirlin:dirlin:prop:2:proof:4}), $\int_0^\infty \phi_{h,i,j}(r) \,dr=\int_0^\infty \phi_{i,j}(r) \,dr\lrp{1+\order{1}}$. Replacing this leads to
\begin{align}
(\ref{kerdirlin:dirlin:prop:2:proof:4})=&\,\frac{c_{h,q}(L)}{ng}\Bigg\{ R(K)\om{q-1}\lrc{\frac{d_q(L)}{\om{q-1}}+\order{1}}f(\bx,z)-R(K)h^2\om{q-1}\lrc{\frac{e_q(L)}{\om{q-1}}+\order{1}}\bx^T\bnab_\bx f(\bx,z)\nonumber\\
&+\frac{R(K)h^4\om{q-1}}{2}\lrc{\frac{1}{\om{q-1}}\frac{\int_0^\infty L^2(r)r^{\frac{q}{2}+1}\,dr}{\int_0^\infty L(r)r^{\frac{q}{2}-1}\,dr}+\order{1}}\bx^T \bHcal_\bx f(\bx,z)\nonumber\\
&+\frac{R(K)h^2\om{q-1}}{2}\lrc{\frac{2e_q(L)}{\om{q-1}}+\order{1}}q^{-1}\lp\nabla^2_\bx f(\bx,z)-\bx^T \bHcal_\bx f(\bx,z)\bx\rp\nonumber\\
&+\frac{\mu_2\lrp{K^2}g^2\om{q-1}}{2}\lrc{\frac{d_q(L)}{\om{q-1}}+\order{1}}\Hcal_z f(\bx,z)\nonumber\\
&+\om{q-1}\lrc{\frac{e_q(L)}{\om{q-1}}+\order{1}}\order{h^2}+\om{q-1}\lrc{\frac{d_q(L)}{\om{q-1}}+\order{1}}\order{g^2}\Bigg\}\nonumber\\
=&\,\frac{c_{h,q}(L)}{ng}\Bigg[R(K)d_q(L)f(\bx,z)+R(K)e_q(L)h^2\Psi_{\bx}f(\bx,z)+\mu_2\lrp{K^2}d_q(L)\frac{g^2}{2}\Hcal_z f(\bx,z)\Bigg]\nonumber\\
&+\order{(nh^qg)^{-1}}.\label{kerdirlin:dirlin:prop:2:proof:5}
\end{align}

The second term of (\ref{kerdirlin:dirlin:prop:2:proof:1}) is 
\begin{align}
\E{\hat f_{h,g}(\bx,z)}^2 %
=\lrc{f(\bx,z)+b_q(L)h^2\Psi_\bx(f,\bx,z)+\frac{g^2}{2}\mu_2(K)\Hcal_z f(\bx,z)}^2+\order{h^2+g^2}.
\label{kerdirlin:dirlin:prop:2:proof:6}
\end{align}

Joining (\ref{kerdirlin:dirlin:prop:2:proof:5}) and (\ref{kerdirlin:dirlin:prop:2:proof:6}),
\begin{align*}
\V{\hat f_{h,g}(\bx,z)}=&\,\frac{c_{h,q}(L)}{ng}\bigg[R(K)d_q(L)f(\bx,z)+R(K)e_q(L)h^2\Psi_{\bx}f(\bx,z)\\
&+\mu_2\lrp{K^2}d_q(L)\frac{g^2}{2}\Hcal_z f(\bx,z)\bigg]\\
&-\frac{1}{n}\lrc{f(\bx,z)+b_q(L)h^2\Psi_\bx(f,\bx,z)+\frac{g^2}{2}\mu_2(K)\Hcal_z f(\bx,z)}^2\\
&+\order{(nh^qg)^{-1}}+\order{n^{-1}(h^2+g^2)},
\end{align*}
which can be simplified into
\begin{align*}
\V{\hat f_{h,g}(\bx,z)}=\frac{c_{h,q}(L)}{ng}R(K)d_q(L)f(\bx,z)+\order{(nh^qg)^{-1}}&.
\end{align*}
\end{proof}


\begin{proof}[Proof of Theorem \ref{kerdirlin:normality:th:1}]
Let $\lrb{\lrp{\bX_i,Z_i}}_{i=1}^n$ be a random sample from the directional-linear random variable $\lrp{\bX,Z}$, whose support is contained in $\Om{q}\times\R$. The directional kernel estimator in a fixed point $(\bx,z)\in\Om{q}\times\R$ can be written as
\begin{align*}
\hat f_{h_n,g_n}(\bx,z)=\frac{1}{n}\sum_{i=1}^n V_{n,i},\quad V_{n,i}=\frac{c_{h,q}(L)}{g} LK\lrp{\frac{1-\bx^T\bX_i}{h_n^2},\frac{z-Z_i}{g_n}},
\end{align*}
where notation $h_n$ and $g_n$ for the bandwidths remarks their dependence on the sample size $n$ given by condition \ref{kerdirlin:cond:dl3}.\\

As $\lrb{\lrp{\bX_i,Z_i}}_{i=1}^n$ is a collection of independent and identically distributed (iid) copies of $(\bX,Z)$, then $\lrb{V_{n,i}}_{i=1}^n$ is also an iid collection of copies of the random variable $V_n=LK\big(\frac{1-\bx^T\bX}{h_n^2},\frac{z-Z}{g_n}\big)$. Then, the Lyapunov's condition ensures that, if for some $\delta>0$ the next condition holds:
\begin{align*}
\lim_{n\to\infty}\frac{\E{\abs{V_n-\E{V_n}}^{2+\delta}}}{n^\frac{\delta}{2}\V{V_n}^{1+\frac{\delta}{2}}}=0,
\end{align*}
then the following central limit theorem is valid:
\begin{align*}
\sqrt{n}\frac{\bar V_{n}-\E{V_n}}{\sqrt{\V{V_n}}}\stackrel{d}{\longrightarrow}\mathcal{N}(0,1),
\end{align*}
where $\bar V_{n}=\frac{1}{n}\sum_{i=1}^n V_{n,i}$. This  condition will be proved for $V_n=LK\big(\frac{1-\bx^T\bX}{h_n^2},\frac{z-Z}{g_n}\big)$.\\

First of all, the order of $\mathbb{E}\big[\abs{V_n}^{2+\delta}\big]$ is
\begin{align*}
\E{\abs{V_n}^{2+\delta}}=&\,\Iqr{\abs{\frac{c_{h_n,q}(L)}{g_n}LK\lrp{\frac{1-\bx^T\by}{h_n^2},\frac{z-t}{g_n}}}^{2+\delta}f(\by,t)}{\by}{t}\\
=&\,\lrp{\frac{c_{h_n,q}(L)}{g_n}}^{2+\delta}\Iqr{LK^{2+\delta}\lrp{\frac{1-\bx^T\by}{h_n^2},\frac{z-t}{g_n}}f(\by,t)}{\by}{t}\\
=&\,\lrp{\frac{c_{h_n,q}(L)}{g_n}}^{2+\delta}g_n h_n^q\int_0^{2h_n^{-2}}\int_{\Om{q-1}}\int_\R LK^{2+\delta}\lrp{r,v}f((\bx,z)+\ba_{\bx,z,\bxi})\,dv\,\om{q-1}(d\bxi)\\
&\times r^{\frac{q}{2}-1}\lrp{2-h_n^2r}^{\frac{q}{2}-1}\,dr\\
=&\,\lrp{\frac{c_{h_n,q}(L)}{g_n}}^{2+\delta}\!\!\!g_n h_n^q\int_0^{2h_n^{-2}}\!\!\!\int_{\Om{q-1}}\!\int_\R LK^{2+\delta}\lrp{r,v}\,dv\,\om{q-1}(d\bxi)\,r^{\frac{q}{2}-1}\lrp{2-h_n^2r}^{\frac{q}{2}-1}\!\!\,dr\\
&\times\lrc{f(\bx,z)+\order{h_n^2+g_n^2}}\\
\sim& \lrp{\frac{c_{h_n,q}(L)}{g_n}}^{2+\delta}g_n h_n^q2^{\frac{q}{2}-1}\om{q-1}\int_0^{\infty}\int_\R LK^{2+\delta}\lrp{r,v}r^{\frac{q}{2}-1}\,dv\,dr \times f(\bx,z)\\
\sim&\lrp{\frac{\lambda_{q}(L)^{-1}h_n^{-q}}{g_n}}^{2+\delta}g_n h_n^q2^{\frac{q}{2}-1}\om{q-1}\int_0^{\infty}\int_\R LK^{2+\delta}\lrp{r,v}r^{\frac{q}{2}-1}\,dv\,dr \times f(\bx,z)\\
=&\,\lrp{h_n^qg_n}^{-(1+\delta)}\times\frac{\int_0^{\infty}\int_\R LK^{2+\delta}\lrp{r,v}r^{\frac{q}{2}-1}\,dv\,dr\times f(\bx,z) }{\lrp{2^{\frac{q}{2}-1}\om{q-1}}^{1+\delta}\lrp{\int_0^\infty L(r)r^{\frac{q}{2}-1}\,dr}^{2+\delta}}\\
=&\,\Order{\lrp{h_n^qg_n}^{-\lrp{1+\delta}}}.
\end{align*}

On the other hand, by Proposition \ref{kerdirlin:dirlin:prop:2}, the variance of $V_n$ has order
\begin{align*}
\V{V_n}=&\,\frac{c_{h,q}(L)}{g_n}R(K)d_q(L)f(\bx,z)+\order{(h_n^qg_n)^{-1}}\sim \frac{R(K)d_q(L)f(\bx,z)}{\lambda_{q}(L)} \frac{1}{h_n^qg_n}=\Order{\lrp{h_n^qg_n}^{-1}}.
\end{align*}

Using that $\mathbb{E}\big[\abs{V_n-\E{V_n}}^{2+\delta}\big]=\mathcal{O}\big(\mathbb{E}\big[\abs{V_n}^{2+\delta}\big]\big)$ (see Remark \ref{kerdirlin:appendix:normality:rem:1}) and by condition \ref{kerdirlin:cond:dl3}, it follows that the Lyapunov's condition is satisfied:
\begin{align*}
\frac{\E{\abs{V_n-\E{V_n}}^{2+\delta}}}{n^\frac{\delta}{2}\V{V_n}^{1+\frac{\delta}{2}}}=\Order{\frac{\lrp{h_n^qg_n}^{-(1+\delta)}}{n^\frac{\delta}{2}\lrp{h_n^qg_n}^{-(1+\frac{\delta}{2})}}}=\Order{\lrp{nh_n^qg_n}^{-\frac{\delta}{2}}}\longrightarrow 0,
\end{align*}
as $n\to\infty$. Therefore,
\begin{align*}
\frac{\hat f_{h_n,g_n}(\bx,z)-\E{\hat f_{h_n,g_n}(\bx,z)}}{\sqrt{\V{\hat f_{h_n,g_n}(\bx,z)}}}\stackrel{d}{\longrightarrow}\mathcal{N}(0,1),
\end{align*}
pointwise for every $(\bx,z)\in\Om{q}\times\R$ (note that $\sqrt{n}$ is included in the variance term). Plugging-in the asymptotic expressions for the bias and the variance results
\begin{align*}
\sqrt{nh_n^qg_n}\lrp{\hat f_{h_n,g_n}(\bx,z)-f(\bx,z)-\mathrm{ABias}\lrc{\hat f_{h_n,g_n}(\bx,z)}}\stackrel{d}{\longrightarrow}\mathcal{N}\lrp{0,R(K)d_q(L)f(\bx,z)}.
\end{align*}
\begin{rem}
	\label{kerdirlin:appendix:normality:rem:1}
	The proof of $\mathbb{E}\big[\abs{V_n-\E{V_n}}^{2+\delta}\big]=\mathcal{O}\big(\mathbb{E}\big[\abs{V_n}^{2+\delta}\big]\big)$ is simple. For example, using the $C_p$ inequality with $p=2+\delta$: $\abs{a+b}^{2+\delta}\leq 2^{1+\delta}\big(\abs{a}^{2+\delta}+\abs{b}^{2+\delta}\big)$, with $a,b\in\mathbb{R}$. Then,
	\begin{align*}
	\E{\abs{V_n-\E{V_n}}^{2+\delta}}&\leq 2^{1+\delta}\E{\abs{V_n}^{2+\delta}+\abs{\E{V_n}}^{2+\delta}}\\
	&=2^{1+\delta}\lrp{\E{\abs{V_n}^{2+\delta}}+\abs{\E{V_n}}^{2+\delta}}\\
	&\leq 2^{2+\delta}\E{\abs{V_n}^{2+\delta}},
	\end{align*}
	where the last step follows by Jensen's inequality applied to the convex function $\abs{\cdot}^{2+\delta}$.
\end{rem}
\end{proof}


\begin{proof}[Proof of Proposition \ref{kerdirlin:dir:prop:3}]
It is straightforward from Proposition \ref{kerdirlin:dir:prop:2} and Lemma \ref{kerdirlin:dir:lem:4}. For a point $\bx$ in $\Om{q}$:
\begin{align*}
\mse{\hat f_h(\bx)}=&\,\lrc{\E{\hat f_h(\bx)}-f(\bx)}^2+\V{\hat f_h(\bx)}\\
=&\,\lrc{b_q(L)\Psi(f,\bx)h^2+\order{h^2}}^2+\frac{c_{h,q}(L)}{n}d_q(L)f(\bx)+\order{(nh^q)^{-1}}\\
=&\,b_q(L)^2\Psi(f,\bx)^2h^4+\frac{c_{h,q}(L)}{n}d_q(L)f(\bx)+\order{h^4+(nh^q)^{-1}}.
\end{align*}
Integrating over $\Om{q}$ in the previous equation,
\begin{align*}
\mise{\hat f_{h}}%
=&\,b_q(L)^2 \Iq{\Psi(f,\bx)^2}{\bx}h^4+\frac{c_{h,q}(L)}{n}d_q(L)+\order{h^4+(nh^q)^{-1}}.
\end{align*}
\end{proof}


\begin{proof}[Proof of Corollary \ref{kerdirlin:dir:cor:1}]
To obtain the bandwidth that minimizes AMISE consider (\ref{kerdirlin:normalizing}) in the previous equation and derive it with respect to $h$:
\begin{align*}
\frac{d}{dh}\amise{\hat f_{h}}=&\,4b_q(L)^2 R\lrp{\Psi(f,\cdot)}h^3-q\lambda_q(L)^{-1}h^{-(q+1)}d_q(L)n^{-1}=0.
\end{align*}
The solution of this equation results in
\begin{align*}
h_{\mathrm{AMISE}}=&\,\lrc{\frac{qd_q(L)}{4b_q(L)^2\lambda_q(L) R(\Psi(f,\cdot)) n}}^{\frac{1}{4+q}}.
\end{align*}
\end{proof}


\begin{proof}[Proof of Proposition \ref{kerdirlin:dirlin:prop:3}] 
It is straightforward from Propositions \ref{kerdirlin:dirlin:prop:1} and \ref{kerdirlin:dirlin:prop:2}:
\begin{align*}
\mse{\hat f_{h,g}(\bx,z)}=&\,\lrc{\E{\hat f_{h,g}(\bx,z)}-f(\bx,z)}^2+\V{\hat f_{h,g}(\bx,z)}\\
=&\,\lrc{h^2b_q(L)\Psi_\bx(f,\bx,z)+\frac{g^2}{2}\Hcal_z f(\bx,z)\mu_2(K)+\order{h^2}+\order{g^2}}^2\\
&+\frac{c_{h,q}(L)}{ng}R(K)d_q(L)f(\bx,z)+\order{(nh^qg)^{-1}}\\
=&\,h^4b_q(L)^2\Psi_\bx(f,\bx,z)^2+\frac{g^4}{4}\mu_2(K)^2\Hcal_z f(\bx,z)^2\\
&+h^2g^2b_q(L)\mu_2(K)\Hcal_z f(\bx,z)\Psi_\bx(f,\bx,z)\\
&+\frac{c_{h,q}(L)}{ng}R(K)d_q(L)f(\bx,z)+\order{h^4+g^4+(nh^qg)^{-1}}.
\end{align*}
Integrating the previous equation and denoting by $I\lrc{\phi}=\Iqr{\phi(\bx,z)}{\bx}{z}$ for a function $\phi:\Omega_q\times\R\rightarrow\R$,
\begin{align*}
\mise{\hat f_{h,g}}=&\,b_q(L)^2I\lrc{\Psi_\bx(f,\cdot,\cdot)^2} h^4+\frac{g^4}{4}\mu_2(K)^2I\lrc{\Hcal_z f(\cdot,\cdot)^2}\\
&+h^2g^2b_q(L)\mu_2(K)I\lrc{\Psi_\bx(f,\cdot,\cdot)\Hcal_z f(\cdot,\cdot)}+\frac{c_{h,q}(L)}{ng}d_q(L)R(K)\\
&+\order{h^4+g^4+(nh^qg)^{-1}}.
\end{align*}
\end{proof}


\begin{proof}[Proof of Corollary \ref{kerdirlin:dirlin:cor:1}]
Suppose that $g=\beta h$ in the previous equation. Again, use that $c_{h,q}(L) \sim \lambda_q(L)^{-1}h^{-q}$ and derive with respect to $h$ to obtain
\begin{align*}
\frac{d}{dh}\amise{\hat f_{h,\beta h}}=4c_1h^3+4c_2h^3+4c_3h^3-(q+1)c_4h^{-(q+2)}=0,
\end{align*}
where
\begin{align*}
c_1=&\,b_q(L)^2I\lrc{\Psi_\bx(f,\cdot,\cdot)^2},\quad c_2=\frac{1}{4}\mu_2(K)^2I\lrc{\Hcal_z f(\cdot,\cdot)^2}\beta^4, \\
c_3=&\,b_q(L)\mu_2(K)I\lrc{\Psi_\bx(f,\cdot,\cdot)\Hcal_z f(\cdot,\cdot)}\beta^2,\quad
c_4=\frac{d_q(L)R(K)}{\lambda_q(L) n\beta}.
\end{align*}
It follows immediately that
\begin{align*}
h_{\mathrm{AMISE}}=&\,\lrc{\frac{(q+1)c_4}{4(c_1+c_2+c_3)}}^{\frac{1}{5+q}}.
\end{align*}
Given that $R\big(b_q(L)\Psi_\bx(f,\cdot,\cdot)+\frac{\beta^2}{2}\mu_2(K)\Hcal_z f(\cdot,\cdot)\big)\allowbreak=c_1+c_2+c_3$, the desired expression is obtained. In the case where $q=1$ it is possible to derive the form of $\beta$ by solving $\frac{\partial}{\partial h}\mathrm{AMISE}\big[\hat f_{h,g}\big]=0$ and $\frac{\partial}{\partial g}\mathrm{AMISE}\big[\hat f_{h,g}\big]=0$. For this case, $\beta$ has the closed form
\[
\beta=\lrp{\frac{\frac{1}{4}\mu_2(K)^2 I\lrc{\Hcal_z f(\cdot,\cdot)^2}}{b_q(L)^2 I\lrc{\Psi_\bx (f,\cdot,\cdot)^2}}}^{\frac{1}{4}}.
\]
\end{proof}


\begin{proof}[Proof of Proposition \ref{kerdirlin:mise:th:1}]
Consider the $r$-mixture of directional von Mises densities given in (\ref{kerdirlin:mise:mvm}). Then:
\begin{align*}
\mise{\hat f_h}=&\,\E{\Iq{\lrp{\hat f_h(\bx)-f_r(\bx)}^2}{\bx}}\\
=&\,\E{\Iq{\hat f_h(\bx)^2-2\hat f_h(\bx)f_r(\bx)+f_r(\bx)^2}{\bx}}\\
=&\,\frac{c_{h,q}(L)^2}{n}\Iq{\Iq{L^2\lrp{\frac{1-\bx^T\by}{h^2}}f_r(\by)}{\bx}}{\by}\\
&+\frac{c_{h,q}(L)^2(n-1)}{n}\int_{\Om{q}}\int_{\Om{q}}\int_{\Om{q}}L\lrp{\frac{1-\bx^T\by}{h^2}}L\lrp{\frac{1-\bx^T\bz}{h^2}}f_r(\by)f_r(\bz)\\
&\times \om{q}(d\bx)\,\om{q}(d\by)\,\om{q}(d\bz)\\
&-2c_{h,q}(L)\Iq{\Iq{L\lrp{\frac{1-\bx^T\by}{h^2}}f_r(\bx)f_r(\by)}{\bx}}{\by}\\
&+\Iq{f_r(\bx)^2}{\bx}\\
=&\,(\co)+(\co)-(\co)+(\co).
\end{align*}

The four terms of the previous equation will be computed separately. The first one is
\addtocounter{equation}{-3}
\begin{align*}
(\arabic{equation})=&\,\frac{c_{h,q}(L)^2}{n}\Iq{\Iq{L^2\lrp{\frac{1-\bx^T\by}{h^2}}f_r(\by)}{\bx}}{\by}\\
=&\,\sum_{j=1}^np_j\frac{c_{h,q}(L)^2}{n}\Iq{\Iq{e^{-2\frac{1-\bx^T\by}{h^2}}C_q(\kappa_j)e^{\kappa_j\by^T\bmu_j}}{\bx}}{\by}\\
=&\,\sum_{j=1}^np_j\frac{c_{h,q}(L)^2}{n}\Iq{\Iq{e^{-\frac{1-\bx^T\by}{\lrp{{h/\sqrt{2}}}^2}}}{\bx}C_q(\kappa_j)e^{\kappa_j\by^T\bmu_j}}{\by}\\
=&\,\sum_{j=1}^np_j\frac{c_{h,q}(L)^2}{c_{h/\sqrt{2},q}(L)n}\Iq{C_q(\kappa_j)e^{\kappa_j\by^T\bmu_j}}{\by}\\
=&\,\sum_{j=1}^np_j\frac{c_{h,q}(L)^2}{c_{h/\sqrt{2},q}(L)n}\\
=&\,\lrp{D_q(h)n}^{-1}.
\end{align*}
The second one is
\addtocounter{equation}{1}
\begin{align*}
(\arabic{equation})=&\,\frac{c_{h,q}(L)^2(n-1)}{n}\Iq{\Iq{\Iq{L\lrp{\frac{1-\bx^T\by}{h^2}}L\lrp{\frac{1-\bx^T\bz}{h^2}}f_r(\by)f_r(\bz)}{\bx}}{\by}}{\bz}\\
=&\,\frac{c_{h,q}(L)^2(n-1)}{n}\int_{\Om{q}}\int_{\Om{q}}\int_{\Om{q}}e^{-2/h^2}e^{\bx^T\by/h^2}e^{\bx^T\bz/h^2}\sum_{j=1}^r\sum_{l=1}^r p_jp_l C_q(\kappa_j)C_q(\kappa_l)e^{\kappa_j\by^T\bmu_j}e^{\kappa_l\bz^T\bmu_l}\\
&\times\,\om{q}(d\bx)\,\om{q}(d\by)\,\om{q}(d\bz)\\
=&\,\frac{c_{h,q}(L)^2(n-1)}{n}e^{-2/h^{2}}\sum_{j=1}^r\sum_{l=1}^r p_jp_l C_q(\kappa_j)C_q(\kappa_l)\\
&\times\Iq{\Iq{\Iq{e^{\bx^T\by/h^{2}}e^{\bx^T\bz/h^{2}}e^{\kappa_j\by^T\bmu_j}e^{\kappa_l\bz^T\bmu_l}}{\bx}}{\by}}{\bz}\\
=&\,\frac{(n-1)}{n}\lrp{(2\pi)^\frac{q+1}{2}h^{q-1}\mathcal{I}_\frac{q-1}{2}(1/h^{2})}^{-1}\sum_{j=1}^r\sum_{l=1}^r p_jp_l C_q(\kappa_j)C_q(\kappa_l)\\
&\times\Iq{\lrc{\Iq{e^{\bx^T\by/h^{2}+\kappa_j\by^T\bmu_j}}{\by}\Iq{e^{\bx^T\bz/h^{2}+\kappa_l\bz^T\bmu_l}}{\bz}}}{\bx}\\
=&\,\big(1-n^{-1}\big)C_q(1/h^2)\sum_{j=1}^r\sum_{l=1}^r p_jp_l C_q(\kappa_j)C_q(\kappa_l)\Iq{\Bigg[\Iq{e^{||\bx/h^2+\kappa_j\bmu_j||\by^T\Big(\frac{\bx/h^2+\kappa_j\bmu_j}{||\bx/h^2+\kappa_j\bmu_j||}\Big)}}{\by}\\
&\times\Iq{e^{||\bx/h^2+\kappa_l\bmu_l||\bz^T\lrp{\frac{\bx/h^2+\kappa_l\bmu_l}{||\bx/h^2+\kappa_l\bmu_l||}}}}{\bz}\Bigg]}{\bx}\\
=&\,\big(1-n^{-1}\big)C_q(1/h^2)\sum_{j=1}^r\sum_{l=1}^r p_jp_l \Iq{\frac{C_q(\kappa_j)C_q(\kappa_l)}{C_q(||\bx/h^2+\kappa_j\bmu_j||)C_q(\norm{\bx/h^2+\kappa_l\bmu_l})}}{\bx}\\
=&\,\big(1-n^{-1}\big)\mathbf{p}^T\mathbf{\Psi_2}(h)\mathbf{p},
\end{align*}
where $\mathbf{\Psi_2}(h)_{r\times r}$ is the matrix with $ij$-th entry $C_q\lrp{1/h^2}\Iq{\frac{C_q(\kappa_j)C_q(\kappa_l)}{C_q(||\bx/h^2+\kappa_j\bmu_j||)C_q(\norm{\bx/h^2+\kappa_l\bmu_l})}}{\bx}$. The third one results in:
\addtocounter{equation}{1}
\begin{align*}
(\arabic{equation})=&\,c_{h,q}(L)\Iq{\Iq{L\lrp{\frac{1-\bx^T\by}{h^2}}f_r(\bx)f_r(\by)}{\bx}}{\by}\\
=&\,c_{h,q}(L)\sum_{j=1}^r\sum_{l=1}^rp_j p_l \Iq{\Iq{e^{-\frac{1-\bx^T\by}{h^2}} C_q(\kappa_j)C_q(\kappa_l)e^{\kappa_j\bx^T\bmu_j}e^{\kappa_l\by^T\bmu_l}}{\bx}}{\by}\\
=&\,c_{h,q}(L)e^{-1/h^{2}}\sum_{j=1}^r\sum_{l=1}^rp_j p_l C_q(\kappa_j)C_q(\kappa_l)\Iq{\Iq{e^{||\by/h^2+\kappa_j\bmu_j||\bx^T\Big(\frac{\by/h^2+\kappa_j\bmu_j}{||\by/h^2+\kappa_j\bmu_j||}\Big)}\\
&\times}{\bx}e^{\kappa_l\by^T\bmu_l}}{\by}\\
=&\,C_q(1/h^2)\sum_{j=1}^r\sum_{l=1}^rp_j p_l C_q(\kappa_j)C_q(\kappa_l)\Iq{\frac{e^{\kappa_l\by^T\bmu_l}}{C_q(||\by/h^2+\kappa_j\bmu_j||)}}{\by}\\
=&\,\mathbf{p}^T\mathbf{\Psi_1}(h)\mathbf{p},
\end{align*}
where the matrix $\mathbf{\Psi_1}(h)_{r\times r}$ has $ij$-th entry  $C_q\lrp{1/h^2}C_q(\kappa_j)C_q(\kappa_l)\Iq{\frac{e^{\kappa_l\by^T\bmu_l}}{C_q(||\by/h^2+\kappa_j\bmu_j||)}}{\by}$. Finally, the fourth term is:
\addtocounter{equation}{1}
\begin{align*}
(\arabic{equation})=&\,\Iq{\bigg(\sum_{j=1}^r p_j f_{\mathrm{vM}}(\bx;\bmu_j,\kappa_j)\bigg)^2}{\bx}\\
=&\,\Iq{\sum_{j=1}^r\sum_{l=1}^r p_jp_l f_{\mathrm{vM}}(\bx;\bmu_j,\kappa_j)f_{\mathrm{vM}}(\bx;\bmu_l,\kappa_l)}{\bx}\\
=&\,\sum_{j=1}^r\sum_{l=1}^r p_jp_lC_q(\kappa_j)C_q(\kappa_l)\Iq{e^{\kappa_j\bx^T\bmu_j}  e^{\kappa_l\bx^T\bmu_l}}{\bx}\\
=&\,\sum_{j=1}^r\sum_{l=1}^r p_jp_lC_q(\kappa_j)C_q(\kappa_l)\Iq{e^{\norm{\kappa_j\bmu_j+\kappa_l\bmu_l}\bx^T\Big(\frac{\kappa_j\bmu_j+\kappa_l\bmu_l}{||\kappa_j\bmu_j+\kappa_l\bmu_l||}\Big)}}{\bx}\\
=&\,\sum_{j=1}^r\sum_{l=1}^r p_jp_l\frac{C_q(\kappa_j)C_q(\kappa_l)}{C_q(||\kappa_j\bmu_j+\kappa_l\bmu_l||)}\\
=&\,\mathbf{p}^T\mathbf{\Psi_0}(h)\mathbf{p},
\end{align*}
where $\mathbf{\Psi_2}(h)_{r\times r}$ represents the matrix with $ij$-th entry $C_q\lrp{1/h^2}\Iq{\frac{C_q(\kappa_j)C_q(\kappa_l)}{C_q(||\by/h^2+\kappa_j\bmu_j||)}}{\by}$. Note that if $\kappa_j\bmu_j+\kappa_l\bmu_l=0$, then $\Iq{}{\bx}=\frac{1}{C_q(0)}=\omega_{q}$ so the result is consistent in this situation.
\end{proof}


\begin{proof}[Proof of Proposition \ref{kerdirlin:mise:th:2}]
Consider the $r$-mixture of directional-linear independent von Mises and normals $f_r(\bx,z)=\sum_{j=1}^r p_j f_{\mathrm{vM}}(\bx;\bmu_j,\kappa_j)\times\phi_{\sigma_j}(z-m_j)$. Hence:
\begin{align}
\mise{\hat f_{h,g}}=&\,\E{\Iqr{\lrp{\hat f_{h,g}(\bx,z)-f_r(\bx,z)}^2}{\bx}{z}}\nonumber\\
=&\,\E{\Iqr{\hat f_{h,g}(\bx,z)^2-2\hat f_{h,g}(\bx,z)f_r(\bx,z)+f_r(\bx,z)^2}{\bx}{z}}\nonumber\\
=&\,\frac{c_{h,q}(L)^2}{ng^2}\Iqr{\Iqr{LK^2\lrp{\frac{1-\bx^T\by}{h^2},\frac{z-t}{g}}f_r(\by,t)}{\bx}{z}}{\by}{t}\nonumber\\
&+\frac{c_{h,q}(L)^2(n-1)}{ng}\int_{\Om{q}\times\R}\int_{\Om{q}\times\R}\int_{\Om{q}\times\R}LK\lrp{\frac{1-\bx^T\by}{h^2},\frac{z-t}{g^2}}\nonumber\\
&\times LK\lrp{\frac{1-\bx^T\bu}{h^2},\frac{z-s}{g}}f_r(\by,t)f_r(\bu,s)\,dz\,\om{q}(d\bx)\,dt\,\om{q}(d\by)\,ds\,\om{q}(d\bu)\nonumber\\
&-2\frac{c_{h,q}(L)}{g}\int_{\Om{q}\times\R}\int_{\Om{q}\times\R}LK\lrp{\frac{1-\bx^T\by}{h^2},\frac{z-t}{g}}f_r(\bx,z)f_r(\by,t)\nonumber\\
&\times \,dz\,\om{q}(d\bx)\,dt\,\om{q}(d\by)\nonumber\\
&+\Iqr{f_r(\bx,z)^2}{\bx}{z}.\label{kerdirlin:mise:th:2:1}
\end{align}
As the directional-kernel is a product kernel and the mixtures are independent the directional and linear parts can be easily disentangled: 
\begin{align}
(\ref{kerdirlin:mise:th:2:1})=&\,n^{-1}\sum_{j=1}^n p_j\lrc{c_{h,q}(L)^2\Iq{\Iq{L^2\lrp{\frac{1-\bx^T\by}{h^2}}f_{\mathrm{vM}}(\by;\bmu_j,\kappa_j)}{\bx}}{\by}}\nonumber\\
&\times\lrc{\frac{1}{g^2}\Ir{\Ir{K^2\lrp{\frac{z-t}{g}}\phi_{\sigma_j}(t-m_j)}{z}}{t}}\nonumber\\
&+\big(1-n^{-1}\big)\sum_{j=1}^n\sum_{l=1}^n p_jp_l\Bigg[c_{h,q}(L)^2\int_{\Om{q}}\int_{\Om{q}}\int_{\Om{q}}L\lrp{\frac{1-\bx^T\by}{h^2}}L\lrp{\frac{1-\bx^T\bu}{h^2}}\nonumber\\
&\times f_{\mathrm{vM}}(\by;\bmu_j,\kappa_j)f_{\mathrm{vM}}(\bu;\bmu_l,\kappa_l)\,\om{q}(d\bx)\,\om{q}(d\by)\,\om{q}(d\bu)\Bigg]\nonumber\\
&\times\lrc{\frac{1}{g}\int_\R\int_\R\int_\R K\lrp{\frac{z-t}{g}}K\lrp{\frac{z-s}{g}}\phi_{\sigma_j}(t-m_j)\phi_{\sigma_l}(s-m_l)\,dz\,dt\,ds}\nonumber\\
&-2\sum_{j=1}^n\sum_{l=1}^n p_jp_l\Bigg[c_{h,q}(L)\int_{\Om{q}}\int_{\Om{q}}L\lrp{\frac{1-\bx^T\by}{h^2}}f_{\mathrm{vM}}(\bx;\bmu_j,\kappa_j)f_{\mathrm{vM}}(\by;\bmu_l,\kappa_l)\nonumber\\
&\times\,\om{q}(d\bx)\,\om{q}(d\by)\Bigg]\times \lrc{\frac{1}{g}\Ir{\Ir{K\lrp{\frac{z-t}{g}}\phi_{\sigma_j}(z-m_j)\phi_{\sigma_l}(z-m_l)}{z}}{t}}\nonumber\\
&+\sum_{j=1}^n\sum_{l=1}^n p_jp_l\lrc{\Ir{\phi_{\sigma_j}\lrp{z-m_j}\phi_{\sigma_l}(z-m_l)}{z}}\times\bigg[\Iq{f_{\mathrm{vM}}(\bx;\bmu_j,\kappa_j)\nonumber\\
&\times f_{\mathrm{vM}}(\bx;\bmu_l,\kappa_l)}{\bx}\bigg].\label{kerdirlin:mise:th:2:2}
\end{align}
The directional parts were calculated in the previous theorem and the linear ones were studied in
\cite{Marron1992} (see also \cite{Wand1995}, page 26). The combination of these two results yields
\begin{align*}
(\ref{kerdirlin:mise:th:2:2})=&\,\lrp{D_q(h)2\pi^\frac{1}{2}ng}^{-1}+\big(1-n^{-1}\big)\mathbf{p}^T\lrc{\mathbf{\Psi_2}(h)\circ\mathbf{\Omega_2}(g)}\mathbf{p}+\mathbf{p}^T\lrc{\mathbf{\Psi_1}(h)\circ\mathbf{\Omega_1}(g)}\mathbf{p}\\
&+\mathbf{p}^T\lrc{\mathbf{\Psi_0}(h)\circ\mathbf{\Omega_0}(g)}\mathbf{p},
\end{align*}
where the $r\times r$ matrices $\mathbf{\Omega_a}(g)$ have the $ij$-th entry equal to $\phi_{\sigma_a}(m_i-m_j)$, $\sigma_a=\big(ag^2+\sigma_i^2+\sigma_j^2\big)^\frac{1}{2}$ for $a=0,1,2$ and $\mathbf{\Psi_a}(h)$ are the matrices of Proposition \ref{kerdirlin:mise:th:1}. The notation $\circ$ denotes the Hadamard product between matrices, \textit{i.e.}, if $(\mathbf{A})_{ij}=a_{ij}$, $(\mathbf{B})_{ij}=b_{ij}$, then $(\mathbf{A}\circ\mathbf{B})_{ij}=a_{ij}b_{ij}$.
\end{proof}


\begin{proof}[Proof of Corollary \ref{kerdirlin:dir:cor:2}]
In virtue of equation (\ref{kerdirlin:kernel_vonmises}), if the kernel of the density estimator (\ref{kerdirlin:kernel_directional}) is $L(r)=e^{-r}$, $r\geq0$, then the kernel estimator is the $n$-mixture of von Mises with means $\bX_i$, $i=1,\ldots,n$, and common concentrations $1/h_P^2$ given by (\ref{kerdirlin:kernel_vonmises}), where $h_P$ is the pilot bandwidth parameter
\end{proof}


\begin{proof}[Proof of Corollary \ref{kerdirlin:dirlin:cor:2}]
It follows immediately from the previous proposition and corollary.
\end{proof}

\section{Proofs of the technical lemmas}
\label{kerdirlin:appendix:proofstechlemmas}

\begin{proof}[Proof of Lemma \ref{kerdirlin:dir:lem:1a}]
Consider the functions
\begin{align*}
\varphi_h(r)&=L(r)r^{\frac{q}{2}-1}(2-h^2r)^{\frac{q}{2}-1}\mathbbm{1}_{[0,2h^{-2})}(r),\\
\varphi(r)&=\lim_{h\to 0} \varphi_h(r)=L(r)r^{\frac{q}{2}-1}2^{\frac{q}{2}-1}\mathbbm{1}_{[0,\infty)}(r).
\end{align*}
Then, proving $\lim_{h\to0}\lambda_{h,q}(L)=\lambda_{q}(L)$ is equivalent to proving $\lim_{h\to0}\int_0^\infty \varphi_h(r) \,dr=\int_0^\infty \varphi(r) \,dr$.\\

Consider first the case $q\geq 2$. As $\frac{q}{2}-1\geq 0$, then $(2-h^2r)^{\frac{q}{2}-1}\leq 2^{\frac{q}{2}-1}$, $\forall h>0$, $\forall r\in[0,2h^{-2})$. Then:
\begin{align*}
\abs{\varphi_h(r)}&\leq L(r)r^{\frac{q}{2}-1}2^{\frac{q}{2}-1}\mathbbm{1}_{[0,2h^{-2})}(r)\leq\varphi(r),\quad \forall r\in [0,\infty), \forall h>0.
\end{align*}
Because $\int_0^\infty \varphi(r) \,dr<\infty$ by condition \ref{kerdirlin:cond:d2} on the kernel $L$, then by the DCT it follows that $\lim_{h\to0}\int_0^\infty \varphi_h(r) \,dr=\int_0^\infty \varphi(r) \,dr$.\\

For the case $q=1$, $\varphi_h(r)=L(r) r^{-\frac{1}{2}} (2-h^2r)^{-\frac{1}{2}}$. Consider now the following decomposition:
\begin{align*}
\int_0^\infty \varphi_h(r) \,dr=\int_0^{\infty} L(r) r^{-\frac{1}{2}}(2-h^2r)^{-\frac{1}{2}}\mathbbm{1}_{[0,h^{-2})}(r) \,dr + \int_{0}^{\infty} L(r) r^{-\frac{1}{2}}(2-h^2r)^{-\frac{1}{2}}\mathbbm{1}_{[h^{-2},2h^{-2})}(r) \,dr.
\end{align*}
The limit of the first integral can be derived analogously with the DCT. As $(2-h^2r)^{-\frac{1}{2}}$ is monotone increasing, then $(2-h^2r)^{-\frac{1}{2}}\leq 1$, $\forall r\in[0,h^{-2})$, $\forall  h>0$. Therefore:
\begin{align*}
\abs{L(r)r^{-\frac{1}{2}}(2-h^2r)^{-\frac{1}{2}}\mathbbm{1}_{[0,h^{-2})}(r)}\leq L(r)r^{-\frac{1}{2}}\mathbbm{1}_{[0,h^{-2})}(r)\leq \varphi(r),\quad \forall r\in[0,\infty),\,\forall h>0.
\end{align*}
Then, as $\lim_{h\to0} L(r) r^{-\frac{1}{2}}(2-h^2r)^{-\frac{1}{2}}\mathbbm{1}_{[0,h^{-2})}(r)=\varphi(r)$ and $\int_0^\infty\varphi(r)\,dr<\infty$ by condition \ref{kerdirlin:cond:d2}, DCT guarantees that $\lim_{h\to0}\int_0^{\infty} L(r) r^{-\frac{1}{2}}(2-h^2r)^{-\frac{1}{2}}\mathbbm{1}_{[0,h^{-2})}(r) \,dr=\int_0^\infty \varphi(r) \,dr$.\\

For the second integral, as a consequence of \ref{kerdirlin:cond:d2} and Remark \ref{kerdirlin:dir:rem:1}, $L$ must decrease faster than any power function. In particular, for some fixed $h_0>0$, $L(r)\leq r^{-1}$, $\forall r\in[h^{-2},2h^{-2})$, $\forall h\in(0,h_0)$. Using this, it results in: 
\begin{align*}
\lim_{h\to 0}\int_{h^{-2}}^{2h^{-2}} L(r) r^{-\frac{1}{2}}(2-h^2r)^{-\frac{1}{2}}\,dr\leq\lim_{h\to 0} \int_{h^{-2}}^{2h^{-2}} r^{-\frac{3}{2}}(2-h^2r)^{-\frac{1}{2}} \,dr=\lim_{h\to 0}h=0.
\end{align*}
This completes the proof.
\end{proof}
\begin{rem}
	\label{kerdirlin:rem:phitcd}
	It is possible to apply the same techniques to prove the result with the functions
	\begin{align*}
	\varphi_{h,i,j,k}(r)&=L^k(r)r^{\frac{q}{2}+i}(2-h^2r)^{\frac{q}{2}-j}\mathbbm{1}_{[0,2h^{-2})}(r),\\
	\varphi_{i,j,k}(r)&=\lim_{h\to 0} \varphi_{h,i,j,k}(r)=L^k(r)r^{\frac{q}{2}+i}2^{\frac{q}{2}-j}\mathbbm{1}_{[0,\infty)}(r),
	\end{align*}
	with $i=-1,0,1$, $j=0,1$ and $k=1,2$. For the cases where $\frac{q}{2}-j\geq 0$, use DCT. For the other cases, subdivide the integral over $[0,2h^{-2})$ into the intervals $[0,h^{-2})$ and $[h^{-2},2h^{-2})$. Then apply DCT in the former and use a suitable power function to make the latter tend to zero in the same way as described previously.
\end{rem}

\begin{proof}[Proof of Lemma \ref{kerdirlin:dir:lem:1}]
Following \cite{Blumenson1960}, if $\bx$ is a vector of norm $r$ with components $x_j$, $j=1,\ldots,n$, with respect to an orthonormal basis in $\R^n$, then the $n$-dimensional spherical coordinates of $\bx$ are given by 
\begin{align}
\label{kerdirlin:dir:lem:1:proof:1}
\lb\begin{array}{l}
\displaystyle x_1=r\cos\phi_1,\\
\displaystyle x_j=r\cos\phi_j\prod_{k=1}^{j-1}\sin\phi_k,\quad j=2,\ldots,n-2,\\
\displaystyle x_{n-1}=r\sin\theta\prod_{k=1}^{n-2}\sin\phi_k,\\
\displaystyle x_n=r\cos\theta\prod_{k=1}^{n-2}\sin\phi_k,
\end{array}\ri\quad J=r^{n-1}\prod_{k=1}^{n-2}\sin^k\phi_{n-1-k}.
\end{align}
where $0\leq\phi_j\leq\pi$, $j=1,\ldots,n-2$, $0\leq\theta<2\pi$ and $0\leq r<\infty$. $J$ denotes the Jacobian of the transformation. Special cases of this parametrization are the polar coordinates ($n=2$),
\begin{align*}
\lb\begin{array}{l}
x_1=r\cos\theta,\\
x_2=r\sin\theta,
\end{array}\ri\quad J=r,
\end{align*}
and the spherical coordinates ($n=3$),
\begin{align*}
\lb\begin{array}{l}
x_1=r\cos\phi,\\
x_2=r\sin\theta\sin\phi,\\
x_3=r\cos\theta\sin\phi,
\end{array}\ri\quad J=r^2\sin\phi.
\end{align*}
Note that sometimes this parametrization appears with the roles of $x_1$ and $x_3$ swapped.\\

To continue with the previous notation, let denote $q=n-1$. Using the spherical coordinates ($r=1$, as the integration is on $\Omega_{n-1}$) and then applying the change of variables
\begin{align}
\label{kerdirlin:dir:lem:1:proof:3}
t=\cos\phi_1,\quad d\phi_1=-(1-t^2)^{-\frac{1}{2}}\,dt,
\end{align}
it follows that
\begin{align*}
\int_{\Omega_{n-1}} f(\bx)&\,\omega_{n-1}(d\bx)\\
=&\,\int_{\Omega_{n-1}} f(x_1,\ldots,x_n)\,d(x_1,\ldots,x_n)\\
\stackrel{\mathclap{(\ref{kerdirlin:dir:lem:1:proof:1})}}{=}\,&\,\,
\int_0^{2\pi}\int_0^\pi\times\stackrel{(n-2)}{\cdots}\times\int_0^\pi f\bigg(\cos\phi_1,\cos\phi_2\sin\phi_1,\ldots,\cos\theta\prod_{k=1}^{n-2}\sin\phi_k\bigg)\\
&\times \prod_{k=1}^{n-2}\sin^k\phi_{n-1-k} \prod_{j=n-2}^1\,d\phi_j \,\,d\theta\\
\stackrel{\mathclap{(\ref{kerdirlin:dir:lem:1:proof:3})}}{=}\,&\,\,\int_0^{2\pi}\int_{-1}^{1}\int_0^\pi\times\stackrel{(n-1)}{\cdots}\times\int_0^\pi f\bigg(t,\cos\phi_2(1-t^2)^{\frac{1}{2}},\ldots,\cos\theta\prod_{k=2}^{n-2}\sin\phi_k(1-t^2)^{\frac{1}{2}}\bigg)\\
&\times \prod_{k=1}^{n-3}\sin^k\phi_{n-1-k}(1-t^2)^\frac{n-2}{2} (1-t^2)^{-\frac{1}{2}}\prod_{j=n-2}^2\,d\phi_j\,dt\,d\theta\\
=&\,\int_{-1}^{1}\int_0^{2\pi}\int_0^\pi\times\stackrel{(n-1)}{\cdots}\times\int_0^\pi  f\bigg(t,\cos\phi_2(1-t^2)^{\frac{1}{2}},\ldots,\cos\theta\prod_{k=2}^{n-2}\sin\phi_k(1-t^2)^{\frac{1}{2}}\bigg)\\
&\times \prod_{k=1}^{n-3}\sin^k\phi_{n-1-k}(1-t^2)^\frac{n-3}{2} \prod_{j=n-2}^2\,d\phi_j \,d\theta \,dt\\
\stackrel{\mathclap{(\ref{kerdirlin:dir:lem:1:proof:1})}}{=}\,&\,\,\int_{-1}^{1}\int_{\Omega_{n-2}} f\lrp{t,(1-t^2)^\frac{1}{2}\xi_1,\ldots,(1-t^2)^\frac{1}{2}\xi_{n-1}} (1-t^2)^\frac{n-3}{2}\\
&\times\,d(\xi_1,\ldots,\xi_{n-1})\,dt\\
=&\, \int_{-1}^{1}\int_{\Omega_{n-2}} f\lrp{t,(1-t^2)^\frac{1}{2}\bxi} (1-t^2)^\frac{n-3}{2}\,\omega_{n-2}(d\bxi) \,dt.
\end{align*}
So, for the $q$-dimensional sphere $\Om{q}$, equation (\ref{kerdirlin:dir:lem:1:2}) follows. Note that as the parametrization (\ref{kerdirlin:dir:lem:1:proof:1}) is invariant to coordinates permutations and $t$ can be placed in any argument of the function. The rest of the arguments will remain having the entries $(1-t^2)^{\frac{n-3}{2}}\bxi$. \\

This expression can be improved using an adequate basis representation. From a fixed point $\by\in\Om{q}$, it is possible to complete an orthonormal basis of $\R^{q+1}$, say $\lrb{\by,\bb_1,\ldots,\bb_{q}}$. So an element $\bx\in\Om{q}$ will be expressed as:
\begin{align*}
\bx=\langle \bx,\by\rangle\by+\sum_{i=1}^q \langle\bx,\bb_i\rangle\bb_i=t\by+(1-t^2)^{\frac{1}{2}}\bxi,
\end{align*}
where $t=\langle\bx,\by\rangle\in[-1,1]$ and $\bxi\in T_\by=\lrb{\boldsymbol\eta\in\Om{q}:\boldsymbol\eta\perp\by}$. %
Related to the basis $\lrb{\by,\bb_1,\ldots,\bb_{q}}$, there are the orthogonal matrix $\bB=\lrp{\by, \bb_1, \ldots, \bb_{q}}_{(q+1)\times(q+1)}$ and the semi-orthogonal matrix $\bB_{\by}=\lrp{\bb_1, \ldots, \bb_{q}}_{(q+1)\times q}$. Using the fact that $\bB$ is an orthonormal matrix, is possible to make the change $\bx=\bB\bz$, with $\det \bB=1$ and $\bB^{-1}\Om{q}=\bB^T\Om{q}=\Om{q}$ (as $\bB$ preserves distances). Then, the relation (\ref{kerdirlin:dir:lem:1:3}) holds:
\begin{align*}
\int_{\Omega_{q}} f(\bx)\,\omega_q(d\bx)=&\,\int_{\bB^{-1}\Omega_{q}} f(\bB\bz) \det{\bB}\,\omega_q(d\bz)\\
=&\,\int_{\Omega_{q}} f(\bB\bz)\,\omega_q(d\bz)\\
\stackrel{\mathclap{(\ref{kerdirlin:dir:lem:1:2})}}{=}\,&\,\,\int_{-1}^{1}\int_{\Omega_{q-1}} f\Big(\bB\big(t,(1-t^2)^\frac{1}{2}\bxi\big)^T\Big) (1-t^2)^{\frac{q}{2}-1}\,\omega_{q-1}(d\bxi)\,dt\\
=&\,\int_{-1}^{1}\int_{\Omega_{q-1}} f\lrp{t\by+(1-t^2)^\frac{1}{2}\bB_{\by}\bxi} (1-t^2)^{\frac{q}{2}-1}\,\omega_{q-1}(d\bxi)\,dt.
\end{align*}
\end{proof}

\begin{proof}[Proof of Lemma \ref{kerdirlin:dir:lem:2}]

Without loss of generality, assume that, by the $q$-spherical coordinates (\ref{kerdirlin:dir:lem:1:proof:1}), $x_i=\cos\phi_1$ and $x_j=\cos\phi_2\sin\phi_1$. Using this, the calculus are straightforward for the integrands $x_i$ and $x_ix_j$ (it is assumed that only the terms with positive index are taken into account in the products):
\begin{align*}
\int_{\Om{q}} x_i\,\om{q}(d\bx)=&\,\int_0^{2\pi}\int_0^\pi\times\stackrel{(q-1)}{\cdots}\times\int_0^\pi \cos\phi_1 \prod_{k=1}^{q-2}\sin^k\phi_{q-k}\sin^{q-1}\phi_1  \prod_{j=q-1}^1\,d\phi_j\,d\theta\\
=&\,\int_0^{2\pi}\int_0^\pi\times\stackrel{(q-2)}{\cdots}\times\int_0^\pi\prod_{k=1}^{q-2}\sin^k\phi_{q-k} \prod_{j=q-1}^2\,d\phi_j\, d\theta \times\int_0^\pi \cos\phi_1\sin^{q-1}\phi_1\,d\phi_1\\
=&\,\omega_{q-1}\times0=0,\\
\int_{\Om{q}} x_ix_j \,\om{q}(d\bx)=&\,\int_0^{2\pi}\int_0^\pi\times\stackrel{(q-1)}{\cdots}\times\int_0^\pi \cos\phi_1 \cos\phi_2\sin\phi_1 \prod_{k=1}^{q-3}\sin^k\phi_{q-k}\sin^{q-2}\phi_2\sin^{q-1}\phi_1\\
&\times \prod_{j=q-1}^1\,d\phi_j \,d\theta\\
=&\,\int_0^{2\pi}\int_0^\pi\times\stackrel{(q-3)}{\cdots}\times\int_0^\pi  \prod_{k=1}^{q-3}\sin^k\phi_{q-k} \prod_{j=q-1}^3\,d\phi_j \,d\theta \\
&\times \int_0^\pi \cos\phi_1\sin^{q}\phi_1  \,d\phi_1 \int_0^\pi \cos\phi_2\sin^{q-2}\phi_2 \,d\phi_2\\
=&\,\omega_{q-2}\times0\times0=0.
\end{align*}
The integrand $x_i^2$ is even simpler, using the fact that the integration is over $\Om{q}$:
\begin{align*}
\int_{\Om{q}} x_i^2\,\om{q}(d\bx)=&\,\frac{1}{q+1}\sum_{k=1}^{q+1} \int_{\Om{q}} x_k^2\,\om{q}(d\bx)=\frac{1}{q+1}\int_{\Om{q}} \sum_{k=1}^{q+1} x_k^2 \,\om{q}(d\bx)=\frac{\om{q}}{q+1}.
\end{align*}
\end{proof}

\begin{proof}[Proof of Lemma \ref{kerdirlin:dir:lem:3}]
For $a=1,2$, $p=0,1$ and $q\geq 1$, the properties of the Gamma function ensure that
\begin{align*}
\int_0^\infty L^a(r) r^{\frac{q}{2}-p}\,dr=\int_0^\infty e^{-ar} r^{\frac{q}{2}-p}\,dr=\frac{\Gamma\lrp{\frac{q}{2}-p+1}}{a^{\frac{q}{2}-p+1}}.%
\end{align*}
Therefore:
\begin{align*}
\lambda_q(L)=&\, 2^{\frac{q}{2}-1} \frac{2\pi^\frac{q}{2}}{\Gamma\lrp{\frac{q}{2}}}\Gamma\lrp{\frac{q}{2}}=\!(2\pi)^\frac{q}{2},\,
b_q(L)=\!\Gamma\lrp{\frac{q}{2}}\frac{q}{2}\bigg/\Gamma\lrp{\frac{q}{2}}\!=\frac{q}{2},\,
d_q(L)=\!\frac{\Gamma\lrp{\frac{q}{2}}}{2^{\frac{q}{2}}}\bigg/\Gamma\lrp{\frac{q}{2}}\!=2^{-\frac{q}{2}}.
\end{align*}
The expression for $c_{h,q}(L)$ arises from the fact that $c_{h,q}(L)=C_q\lrp{1/h^2}e^{1/h^2}$.
\end{proof}

\begin{proof}[Proof of Lemma \ref{kerdirlin:dir:lem:4}]
This proof is a rebuild of the one given in \cite{Zhao2001} and is included for the aim of completeness of this work. Furthermore, many techniques used in this proof are also helpful for the proofs of other results in this paper.\\

Let denote $\mathrm{Bias}\big[\hat f_h(\bx)\big]=\mathbb{E}\big[\hat f_h(\bx)\big]-f(\bx)$. To compute the bias, use Lemma \ref{kerdirlin:dir:lem:1} for the change of variables with the orthonormal and semi-orthonormal matrices $\bB=(\bx,\bb_1,\ldots,\bb_{q})$ and $\bB_{\bx}=(\bb_1,\ldots,\bb_{q})$, and then apply the ordinary change of variables
\begin{align}
r=\frac{1-t}{h^2},\quad dr=-h^{-2}\,dt. \label{kerdirlin:dir:prop:1:proof:1a}
\end{align}
This results in:
\begin{align}
\mathrm{Bias}\lrc{\hat f_h(\bx)}=&\,c_{h,q}(L)\E{L\lrp{\frac{1-\bx^T\bX}{h^2}}}-f(\bx)\nonumber\\
=&\,c_{h,q}(L)\int_{\Omega_q}L\lrp{\frac{1-\bx^T\by}{h^2}}f(\by)\,\om{q}(d\by)-c_{h,q}(L)\int_{\Omega_q}L\lrp{\frac{1-\bx^T\by}{h^2}}\,\om{q}(d\by)f(\bx)\nonumber\\
=&\,c_{h,q}(L)\int_{\Omega_q}L\lrp{\frac{1-\bx^T\by}{h^2}}\lrp{f(\by)-f(\bx)}\,\om{q}(d\by)\nonumber\\
=&\,c_{h,q}(L)\int_{-1}^1\int_{\Omega_{q-1}} L\lrp{\frac{1-t}{h^2}} \lrp{f\lrp{t\bx+(1-t^2)^{\frac{1}{2}}\bB_{\bx}\bxi}-f(\bx)}\nonumber\\
&\times (1-t^2)^{\frac{q}{2}-1}\,\omega_{q-1}(d\bxi)\,dt\nonumber\\
\stackrel{\mathclap{(\ref{kerdirlin:dir:prop:1:proof:1a})}}{=}\,&\,\,c_{h,q}(L)h^q\int_{0}^{2h^{-2}}\int_{\Omega_{q-1}} L(r) \lrp{f\lrp{\bx+\ba_{\bx,\bxi}}-f(\bx)}r^{\frac{q}{2}-1} (2-h^2r)^{\frac{q}{2}-1}\,\omega_{q-1}(d\bxi) \,dr\nonumber\\
=&\,c_{h,q}(L)h^q\int_{0}^{2h^{-2}}L(r)r^{\frac{q}{2}-1}(2-h^2r)^{\frac{q}{2}-1}\int_{\Omega_{q-1}}  \lrp{f\lrp{\bx+\ba_{\bx,\bxi}}-f(\bx)}\nonumber\\
&\times \,\omega_{q-1}(d\bxi) \,dr,\label{kerdirlin:dir:prop:1:proof:1}
\end{align}
where $\ba_{\bx,\bxi}=-rh^2\bx+h\lrc{r(2-h^2r)}^{\frac{1}{2}}\bB_{\bx}\bxi\in\Om{q}$. By condition \ref{kerdirlin:cond:d1}, the Taylor expansion of $f$ at $\bx$\nolinebreak[4] is
\begin{align*}
f(\bx+\ba_{\bx,\bxi})-f(\bx)=&\,\ba_{\bx,\bxi}^T\bnab f(\bx)+\frac{1}{2}\ba_{\bx,\bxi}^T\bHcal f(\bx)\ba_{\bx,\bxi}+\order{\ba_{\bx,\bxi}^T\ba_{\bx,\bxi}},
\end{align*}
so the calculus of (\ref{kerdirlin:dir:prop:1:proof:1}) can be split in three parts. For the first use that the integration of $\xi_i$ vanishes by Lemma \ref{kerdirlin:dir:lem:2}:
\begin{align}
\int_{\Om{q-1}} \ba_{\bx,\bxi}^T\bnab f(\bx)\,\om{q-1}(d\bxi)=&\,-rh^2\int_{\Om{q-1}} \bx^T\bnab f(\bx)\,\om{q-1}(d\bxi)\nonumber\\
&+h\lrc{r(2-h^2r)}^{\frac{1}{2}}\int_{\Om{q-1}} \bxi^T \bB_{\bx}^T\bnab f(\bx)\,\om{q-1}(d\bxi)\nonumber\\
=&\,-rh^2 \om{q-1} \bx^T\bnab f(\bx)\label{kerdirlin:dir:prop:1:proof:2}
\end{align}

In the second, by the results of Lemma \ref{kerdirlin:dir:lem:2},
\begin{align}
\int_{\Om{q-1}} \ba_{\bx,\bxi}^T\bHcal f(\bx)\ba_{\bx,\bxi} \,\om{q-1}(d\bxi)=&\,r^2h^4\int_{\Om{q-1}} \bx^T\bHcal f(\bx)\bx\,\om{q-1}(d\bxi)\nonumber\\
&-2rh^3\lrc{r(2-h^2r)}^{\frac{1}{2}}\int_{\Om{q-1}} \bx^T\bHcal f(\bx)\bB_{\bx}\bxi\,\om{q-1}(d\bxi)\nonumber\\
&+h^2 r(2-h^2r)\int_{\Om{q-1}} \bxi^T \bB_{\bx}^T\bHcal f(\bx)\bB_{\bx}\bxi\,\om{q-1}(d\bxi)\nonumber\\
=&\,r^2h^4\om{q-1} \bx^T\bHcal f(\bx)\bx\nonumber\\
&+h^2 r(2-h^2r)\int_{\Om{q-1}}\sum_{i,j=1}^{q}  \bb_i^T\bHcal f(\bx)\bb_j\xi_i\xi_j\,\om{q-1}(d\bxi)\nonumber\\
=&\,r^2h^4\om{q-1} \bx^T\bHcal f(\bx)\bx\nonumber\\
&+h^2 r(2-h^2r)\sum_{i=1}^q  \bb_i^T\bHcal f(\bx)\bb_i\int_{\Om{q-1}}\xi_i^2 \,\om{q-1}(d\bxi)\nonumber\\
=&\,r^2h^4\om{q-1} \bx^T\bHcal f(\bx)\bx\nonumber\\
&+h^2 r(2-h^2r)\om{q-1}q^{-1}\lrc{\nabla^2 f(\bx)-\bx^T \bHcal f(\bx)\bx}.\label{kerdirlin:dir:prop:1:proof:3}
\end{align}
In the last step it is used that by $\sum_{i=1}^q \bb_i\bb_i^T+\bx\bx^T=\bB_{\bx}\bB_{\bx}^T=\bI_{q+1}-\bx\bx^T$,
\begin{align*}
\sum_{i=1}^q \bb_i^T \bHcal f(\bx) \bb_i=\tr{\bHcal f(\bx) \sum_{i=1}^q \bb_i\bb_i^T}=\tr{\bHcal f(\bx)\lrp{\bI_{q+1}-\bx\bx^T}}=\nabla^2 f(\bx)-\bx^T\bHcal f(\bx) \bx.
\end{align*}

Apart from this, the order of the Taylor expansion is
\begin{align}
\order{\ba_{\bx,\bxi}^T\ba_{\bx,\bxi}}=\order{r^2h^4+h^2r(2-h^2r)}=\order{r^2h^4+2h^2r-h^4r^2}=r\order{h^2}. \label{kerdirlin:dir:prop:1:proof:4}
\end{align}
Adding  (\ref{kerdirlin:dir:prop:1:proof:2})--(\ref{kerdirlin:dir:prop:1:proof:4}),
\begin{align}
(\ref{kerdirlin:dir:prop:1:proof:1})=&\,\om{q-1}c_{h,q}(L)h^q\int_{0}^{2h^{-2}}L(r)r^{\frac{q}{2}-1}(2-h^2r)^{\frac{q}{2}-1}\Bigg\{-rh^2 \bx^T\bnab f(\bx)+\frac{r^2h^4}{2} \bx^T\bHcal f(\bx)\bx\nonumber\\
&+ \frac{h^2 r(2-h^2r)}{2q}\lrp{\nabla^2 f(\bx)-\bx^T \bHcal f(\bx)\bx}+r\,\order{h^2}\Bigg\} \,dr\nonumber\\
=&\,-h^2\om{q-1}\lrc{\int_{0}^{2h^{-2}}c_{h,q}(L)h^{q}L(r)r^{\frac{q}{2}}(2-h^2r)^{\frac{q}{2}-1}\,dr}\bx^T\bnab f(\bx)\nonumber\\
&+\frac{h^4\om{q-1}}{2}\lrc{\int_{0}^{2h^{-2}}c_{h,q}(L)h^{q}L(r)r^{\frac{q}{2}+1}(2-h^2r)^{\frac{q}{2}-1}\,dr}\bx^T \bHcal f(\bx) \bx\nonumber\\
&+\frac{h^2\om{q-1}}{2}\lrc{\int_{0}^{2h^{-2}}c_{h,q}(L)h^{q}L(r)r^{\frac{q}{2}}(2-h^2r)^{\frac{q}{2}}\,dr}q^{-1}\lrp{\nabla^2 f(\bx)-\bx^T\bHcal f(\bx) \bx}\nonumber\\
&+\om{q-1}\lrc{\int_{0}^{2h^{-2}}c_{h,q}(L)h^qL(r)r^{\frac{q}{2}}(2-h^2r)^{\frac{q}{2}-1}\,dr}\order{h^2}.\label{kerdirlin:dir:prop:1:proof:5}
\end{align}

Consider the following functions for $h>0$ and $i,j=0,1$:
\begin{align*}
\varphi_{h,i,j}(r)=c_{h,q}(L)h^{q}L(r) r^{\frac{q}{2}+i}(2-h^2r)^{\frac{q}{2}-j}\mathbbm{1}_{[0,2h^{-2})}(r),\quad r\in[0,\infty).
\end{align*}
When $n\to\infty$, $h\to0$ and the limit of $\varphi_{h,i,j}$ is given by
\begin{align*}
\varphi_{i,j}(r)=\lim_{h\to0}\varphi_{h,i,j}(r)=\lambda_q(L)^{-1}L(r) r^{\frac{q}{2}+i} 2^{\frac{q}{2}-j}\mathbbm{1}_{[0,\infty)}(r).
\end{align*}
Then, by Remark \ref{kerdirlin:rem:phitcd} and Lemma \ref{kerdirlin:dir:lem:1a}:
\begin{align*}
\lim_{h\to0}\int_0^\infty \varphi_h(r) \,dr=\lambda_q(L)^{-1}2^{\frac{q}{2}-j}\int_0^\infty L(r) r^{\frac{q}{2}-i}\,dr\stackrel{(\ref{kerdirlin:dir:lem:1a:1})}{=}\lb\begin{array}{ll}
\frac{2^{1-j}}{\om{q-1}}b_q(L),&i=0,\\
\frac{2^{1-j}}{\om{q-1}}\frac{\int_0^\infty L(r)r^{\frac{q}{2}+1}\,dr}{\int_0^\infty L(r)r^{\frac{q}{2}-1}\,dr},&i=1.\\
\end{array}\ri
\end{align*}

So, for the terms between square brackets of (\ref{kerdirlin:dir:prop:1:proof:5}), $\int_0^\infty \varphi_h(r) \,dr=\int_0^\infty \varphi(r) \,dr\lrp{1+\order{1}}$. Replacing this in (\ref{kerdirlin:dir:prop:1:proof:5}) leads to
\begin{align*}
(\ref{kerdirlin:dir:prop:1:proof:5})=&\,-h^2\om{q-1}\lrc{\frac{b_q(L)}{\om{q-1}}+\order{1}}\bx^T\bnab f(\bx)\nonumber\\
&+\frac{h^4\om{q-1}}{2}\lrc{\frac{b_q(L)}{\om{q-1}}\frac{\int_0^\infty L(r)r^{\frac{q}{2}+1}\,dr}{\int_0^\infty L(r)r^\frac{q}{2}\,dr}+\order{1}} \bx^T\bHcal f(\bx)\bx\nonumber\\
&+\frac{h^2\om{q-1}}{2}\lrc{\frac{b_q(L)}{\om{q-1}}+\order{1}} q^{-1}\lrp{\nabla^2 f(\bx)-\bx^T\bHcal f(\bx) \bx}+\om{q-1}\lrc{\frac{b_q(L)}{\om{q-1}}+\order{1}}\order{h^2}\nonumber\\
=&\,h^2b_q(L)\lrc{-\bx^T\bnab f(\bx)+q^{-1}\lrp{\nabla^2 f(\bx)-\bx^T\bHcal f(\bx) \bx}}+\Order{h^4}+\order{h^2}\nonumber\\
=&\,h^2b_q(L)\Psi(f,\bx)+\order{h^2}.
\end{align*}
\end{proof}

\end{document}